\documentclass[useAMS,usenatbib]{mn2e}
\usepackage{graphicx}
\usepackage{latexsym}
\usepackage{psfig}
\usepackage{amssymb,amsmath}
\usepackage{subfig}
\usepackage{caption}

\title{Massive stars in massive clusters II: Disruption of bound clusters by photoionization}

\author[J. E. Dale, B. Ercolano, I.A. Bonnell]{J. E. Dale$^{1}$\thanks{E-mail: dale@usm.lmu.de (JED)}, B. Ercolano$^{1}$, I. A. Bonnell$^{2}$\\
$^{1}$Excellence Cluster `Universe', Boltzmannstr. 2, 85748 Garching, Germany.\\
$^{2}$Department of Physics and Astronomy, University of St Andrews, North Haugh, St Andrews, Fife KY16 9SS}

\begin{document}

\pagerange{\pageref{firstpage}--\pageref{lastpage}} \pubyear{2006}

\maketitle

\label{firstpage}

\def\mnras{MNRAS}
\def\apj{ApJ}
\def\aj{AJ}
\def\aap{A\&A}
\def\apjl{ApJL}
\def\apjs{ApJS}
\def\araa{ARA\&A}
 
\begin{abstract}
We present an SPH parameter study of the dynamical effect of photoionization from O--type stars on star--forming clouds of a range of masses and sizes during the time window before supernovae explode. Our model clouds all have the same degree of turbulent support initially, the ratio of turbulent kinetic energy to gravitational potential energy being set to $E_{\rm kin}/|E_{\rm pot}|$=0.7. We allow the clouds to form stars and study the dynamical effects of the ionizing radiation from the massive stars or clusters born within them. We find that dense filamentary structures and accretion flows limit the quantities of gas that can be ionized, particularly in the higher density clusters. More importantly, the higher escape velocities in our more massive (10$^{6}$M$_{\odot}$) clouds prevent the HII regions from sweeping up and expelling significant quantities of gas, so that the most massive clouds are largely dynamically unaffected by ionizing feedback. However, feedback has a profound effect on the lower--density 10$^{4}$ and 10$^{5}$M$_{\odot}$ clouds in our study, creating vast evacuated bubbles and expelling tens of percent of the neutral gas in the 3Myr timescale before the first supernovae are expected to detonate, resulting in clouds highly porous to both photons and supernova ejecta.
\end{abstract}

 
\section{Introduction}
The vast majority of stars form in giant molecular clouds (GMCs) as members of embedded clusters \citep{2003ARA&A..41...57L}. By comparing birthrates of embedded clusters to the number of surviving open clusters, \cite{2003ARA&A..41...57L} concluded that up to 90$\%$ of clusters are disrupted during their embedded phase. It has long been thought that GMCs and embedded clusters are initially gravitationally bound \citep[e.g.][]{1979IAUS...84...35S}. These two observations can be reconciled if gas ejection by stellar feedback while the gas:stars mass ratio is large unbinds clouds and clusters by decreasing the gravitational potential too quickly for the clusters to adjust. \cite{1980ApJ...235..986H} and later \cite{1997MNRAS.284..785G}, \cite{2003MNRAS.338..673B}, \cite{2003MNRAS.338..665B} and \cite{2006MNRAS.373..752G} studied the effects of removing residual gas from embedded clusters at various rates. They showed that lower star formation efficiencies and shorter gas removal timescales result in the unbinding of larger fractions of stars, up to the total disruption of clusters -- a process often referred to as `infant mortality'. However, none of these authors modelled the gas removal process itself.\\
\indent There are several feedback mechanisms which are thought to be able to influence the dynamics of whole molecular clouds such as HII regions, winds, jets, and supernovae. By considering the momentum injected by these various mechanisms, \cite{2002ApJ...566..302M} concluded that HII region expansion was the most important for clouds with masses $\gtrsim10^{5}$M$_{\odot}$. Analytical and numerical models of ionization--driven champagne flows \citep{1979A&A....71...59T,1979ApJ...233...85B,1979MNRAS.186...59W,1997ApJ...476..166W} have shown that they could be an efficient dispersal mechanism of uniform clouds if the massive stars were located near the peripheries of the clouds. In contrast, work by \cite{1980A&A....90...65M} and \cite{1989A&A...216..207Y} showed that the dispersal efficiency was strongly reduced by the action of gravity or by placing the stars deep inside the clouds. GMCs have very complex density structures and turbulent velocity fields and clusters are usually found deeply embedded inside them. The massive stars in turn are usually to be found near cluster centres, either because they formed there through competitive accretion \citep[e.g][]{2004MNRAS.349..735B} or migrated there through rapid mass segregation \citep[e.g][]{2007ApJ...655L..45M,2009ApJ...700L..99A,2009MNRAS.396.1864M}.\\
\indent Several authors have modelled ionizing feedback from embedded massive stars on their parent molecular clouds. One--dimensional models of the evolution of 2$\times10^{5}$-- 5$\times10^{6}$M$_{\odot}$ GMCs by \cite{2006ApJ...653..361K} concluded that HII regions would destroy clouds $\gtrsim10^{6}$M$_{\odot}$ on timescales of $\sim30$Myr, corresponding to a few freefall times, whilst the lower--mass clouds would survive only $\sim10$Myr. \cite{2011ApJ...738..101G} constructed semianalytic models in which they examined the energy input in a GMC from both internal sources (HII regions) and external sources (accretion of additional mass) and found that these two sources were of roughly equal importance in driving turbulence within the clouds. \cite{2005MNRAS.358..291D} and \cite{2011MNRAS.414..321D} found that accretion flows onto the O--stars strongly limit the effect of radiation by only allowing it to escape into a small fraction of the sky as seen from each source. \cite{2010ApJ...719..831P} also found that the expansion of HII regions could be limited by hydrodynamic flows. \cite{2011arXiv1109.3478W} investigated the evolution of fractal molecular clouds with central ionizing sources. They show that the gas distribution prior to the ignition of the ionizing sources is of crucial importance in determining the outcome of the simulations, and that the principle effect of ionization is to enhance density contrasts that were already present. \cite{2010ApJ...715.1302V} examine the effect of ionizing feedback on both accretion of material onto GMCs themselves, and on the star formation rates within. They find that feedback generally reduces the star formation efficiency, but that the effect is smaller for more massive and denser clouds.\\
\indent The works cited above have only examined a limited portion of the parameter space of possible GMCs. In this paper, we begin a parameter study where we construct undriven turbulent molecular clouds with a variety of masses and radii, allow them to form stars, and model the effects of the photoionizing feedback from their stellar populations. \cite{2009ApJ...699.1092H} have re--examined the data from \cite{1987ApJ...319..730S} and produced a catalogue of masses, radii and velocity dispersions for 158 clouds which we use to define the mass--radius parameter space for this study. As pointed out by \cite{2011MNRAS.413.2935D}, most of the clouds in \cite{2009ApJ...699.1092H}'s catalogue are in fact not gravitationally bound, in contradiction to the common assumption about star--forming GMCs. However, the virial ratio of the clouds in the \cite{2009ApJ...699.1092H} sample is not independent of cloud mass, but instead declines with increasing mass, so that more massive clouds are more likely to be bound, and in fact very few clouds with masses in excess of $10^{4}$M$_{\odot}$ were found to be unbound by \cite{2011MNRAS.413.2935D}. Since the cloud mass function is rather shallow, with a power--law slope of $\sim$-1.6-- -1.8, the massive objects, although less numerous, contain most of the total mass. Unbound clouds may be common by number, but most mass resides in clouds that are bound. Therefore, in this study we restrict ourselves to studying clouds where the ratio of turbulent kinetic to potential energy is less than unity. To begin with, we reduce the size of the parameter space by insisting that all clouds have the same degree of initial turbulent support with the ratio $E_{\rm kin}/|E_{\rm pot}|$ set to 0.7, chosen so that each cloud would be roughly midway between virial equilibrium and marginal boundedness if the turbulent kinetic energy were the only means of support. Our aim in this is to answer the question of whether ionizing feedback can in principle disrupt bound molecular clouds and terminate star formation inside them. We will extend the parameter space to include clouds in which the turbulent kinetic energy exceeds the potential energy in later work.\\
\indent We describe our numerical techniques in Section 2, present our results in Section 3 and our discussion and conclusions follow in Sections 4 and 5 respectively.\\
\section{Numerical methods}
\begin{figure}
\includegraphics[width=0.5\textwidth]{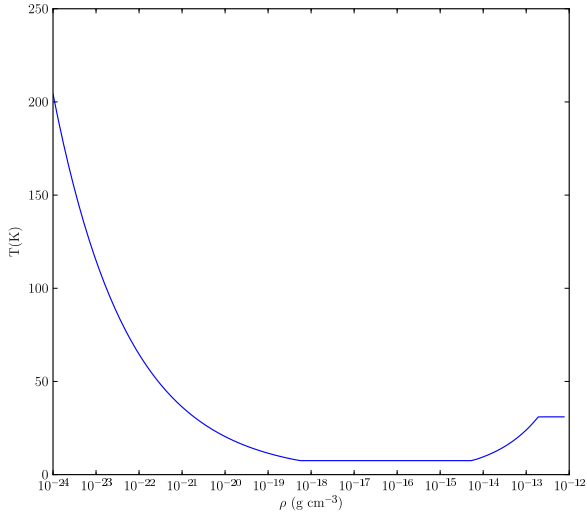}
\caption{Relation between density and temperature used in our calculations.} 
\label{fig:rho_T}
\end{figure}

We use a well--known variant of the Benz \citep{1990nmns.work..269B} Smoothed Particle Hydrodynamics \citep{1992ARA&A..30..543M} code, which is ideal for studying the evolution of molecular clouds and embedded clusters. In all our simulations, we begin with 10$^{6}$ gas particles.  We use the standard artificial viscosity prescription, with $\alpha=1$, $\beta=2$. Particles are evolved on individual timesteps. The code is a hybrid N--body SPH code in which stars are represented by point--mass sink particles \citep{1995MNRAS.277..362B}. Self--gravitational forces between gas particles are calculated using a binary tree, whereas gravitational forces involving sink--particles are computed by direct summation. Sink particles are formed dynamically and may accrete gas particles and grow in mass. In our simulations of 10$^{5}$ and 10$^{6}$M$_{\odot}$ clouds, the sink particles represent stellar clusters, since the mass resolution is not sufficient to capture individual stars. The accretion radii of the clusters are chosen to be 0.25 pc in our Runs A, B, X and D and 0.1pc in Runs E and F, so that the accretion radii of the sinks (our effective cluster radii) are always $\lesssim$1$\%$ of the radius of the simulated clouds. Clusters approaching each other to within their accretion radii are merged if they are mutually gravitationally bound. In our 10$^{4}$M$_{\odot}$ simulations, sink particles represent individual stars. Their accretion radii are set to 0.005pc ($\sim10^{3}$ AU) and mergers are not permitted. In all simulations gravitational interactions of sink particles with other sink particles are smoothed within their accretion radii.\\
\indent We treat the thermodynamics of the neutral gas using a piecewise barotropic equation of state from \cite{2005MNRAS.359..211L}. The use of an equation of state is of course an approximation in lieu of either using prescribed heating and cooling functions or attempting extremely expensive full radiative transfer calculations, but the one we employ here broadly reproduces the findings of more sophisticated treatments of ISM themodynamics. The equation of state we have chosen was originally conceived by \cite{1985MNRAS.214..379L} based on observations of cloud temperatures and densities collected by \cite{1978ApJ...225..380M} and was intended in particular to capture the increase in the gas cooling rate with increasing density at low densities, so that $T\sim\rho^{-0.3}$, until the gas becomes thermally coupled to the dust at a density of $\sim10^{-19}$ g cm$^{-3}$ . This equation of state has proved to be a robust approximation in both atomic and molecular gas and has been recovered by several authors using either cooling and heating functions or performing radiative transfer calculations. \cite{2000ApJ...532..980K} obtained a very similar temperature--density relation at densities below $\sim2\times10^{-20}$g cm$^{-3}$ based on thermal equilibrium calculations from \cite{1995ApJ...443..152W}. \cite{2007ApJ...659.1317G}, \cite{2010MNRAS.404....2G} and \cite{2012MNRAS.tmp.2154G} all recovered very similar temperature--density relations in studies of molecular cloud and star formation.\\
\indent Our equation of state is defined so that $P = k \rho^{\gamma}$, where
\begin{eqnarray}
\begin{array}{rlrl}
\gamma  &=  0.75  ; & \hfill &\rho \le \rho_1 \\
\gamma  &=  1.0  ; & \rho_1 \le & \rho  \le \rho_2 \\
\gamma  &=  1.4  ; & \hfill \rho_2 \le &\rho \le \rho_3 \\
\gamma  &=  1.0  ; & \hfill &\rho \ge \rho_3, \\
\end{array}
\label{eqn:eos}
\end{eqnarray}
and $\rho_1= 5.5 \times 10^{-19} {\rm g\ cm}^{-3} , \rho_2=5.5 \times10^{-15} {\rm g cm}^{-3} , \rho_3=2 \times 10^{-13} {\rm g\ cm}^{-3}$. At low densities, $\gamma$ is less than unity, which mimics the effects of line cooling and implicitly heats very low density gas above the canonical molecular gas temperature of $\sim10$K. The isothermal $\gamma=1.0$ segment at moderate densities approximates the effect of dust cooling and the $\gamma=1.4$ segment represents the regime where dense collapsing cores become optically thick and behave adiabatically. The final isothermal phase of the equation of state is simply in order to allow sink-particle formation to occur. Once the minimum gas temperature, which we set to 7.5K, is specified, the relation between $\rho$ and $T$, which we plot in Figure \ref{fig:rho_T}, is fixed. All our simulated clouds have initial average densities $< \rho_{1}$, so that they lie initially in the line--cooling--dominated regime.\\
\indent We use the photoionization code described in \cite{2007MNRAS.382.1759D} and \cite{2011MNRAS.414..321D}. The code uses a simple ray--tracing algorithm and a Str\"omgren volume technique to compute the flux of ionizing photons arriving at a given SPH particle and update its ionization state accordingly. The on--the--spot approximation is used and the modified recombination coefficient $\alpha_{\rm B}$ is taken to be $3.0\times10^{-13}$ cm$^{3}$ s$^{-1}$. Fully ionized particles are given a temperature of 10$^{4}$ K, whereas partially ionized particles are given temperatures computed from multiplying their ionization fraction by 10$^{4}$ K. Ionised particles that are deprived of photons are allowed to recombine on their individual recombination timescales. If their ionization fractions fall below 0.1, they are considered to be fully neutral once more and they are then allowed to descend the cooling curve from \cite{1993A&A...273..318S}.\\
\indent The ionization algorithm was modified in a simple way in \cite{2011MNRAS.414..321D} to cope with the action of multiple ionizing sources with overlapping HII regions -- we assumed in that work that if a given ionized particle of number density $n$ was receiving photons from $N$ sources, then the volume recombination rate as seen by each source was $\alpha_{\rm B}n^{2}/N$. This approximation is rather crude and we have updated the algorithm to make it more physically realistic. In truth, the number of photons subtracted from the beams from each source passing though a multiply--illuminated particle is proportional to the photon flux from each source as a fraction of the total flux passing through the particle. If an ionized particle is being illuminated by $i$ sources and receives a photon flux $F_{\rm i}$ from each one, the volume recombination rate seen by each source is then $\alpha_{\rm B} n^{2}F_{\rm i}/\sum F_{\rm i}$. We have implemented this in a new version of our algorithm, which iterates, recomputing the individual and total fluxes received by all particles each time until the number of ionized particles converges to an accuracy of 0.1$\%$. We will describe this technique in detail in Dale and Ercolano (2012), in prep, and validate it by comparison with the Monte Carlo radiative transfer code {\sc mocassin} \citep{2003MNRAS.340.1136E,2005MNRAS.362.1038E,2008ApJS..175..534E}. As described below, we repeat the simulation from \cite{2011MNRAS.414..321D} using the new algorithm and compare the results, finding only minor differences in the dynamical influence of ionization.\\
\indent In some of our clouds, our mass resolution is sufficient to regard sink particles as individual stars. We assign each star an ionizing photon flux dependent on its mass M$_{*}$. To reduce the number of low--mass, low--flux ionizing sources which are likely to slow down simulations but contribute few photons, we neglect ionizing sources with masses less than 20M$_{\odot}$. There is a knee in the relation between mass and ionizing luminosity at this mass, so that the total photon budget of a given stellar system will be dominated by stars whose masses exceed 20M$_{\odot}$. We assign photon fluxes according to the formula
\begin{eqnarray}
{\rm log}(Q_{\rm H})=48.1+0.02(M_{*}-20M_{\odot}),
\end{eqnarray}
an approximate fit to the ionizing photon fluxes of solar--metallicity stars tabulated in \cite{1998ApJ...501..192D}.\\
\indent In our simulations of higher mass clouds, our mass resolution is insufficient to follow the formation of individual stars and we instead treat the sink particles as small clusters. We use the criterion from \cite{2011MNRAS.414..321D} to determine their photon fluxes as follows. We compute, assuming a Salpeter mass function between $0.1$ and 100M$_{\odot}$, the mass in stars of more than 30M$_{\odot}$ and divide this quantity by 30M$_{\odot}$ (assuming that such a star is a typical O--star). We then multiply this by the photon flux appropriate for this mass from the above formula, $\sim2\times10^{48}$ s$^{-1}$ (not 10$^{49}$ s$^{-1}$ as incorrectly given in \cite{2011MNRAS.414..321D}). This is evidently a very crude means of estimating the subclusters' luminosities but any such estimate will be crude, since the form and limits of the mass function must be assumed. In addition, except for rather massive clusters, the high--mass end of the stellar mass function will be poorly sampled and the actual numbers and masses of O--stars are therefore very uncertain. We subsequently find, as shown in the Appendix, that uncertainties of a factor a few in the ionizing luminosities of our sources have very little influence on our results.\\
\indent Our model clouds initially have a Gaussian three--dimensional density profile. We seed the gas with a Kolmogorov turbulent velocity field whose total kinetic energy is equal in magnitude to 7/10 the cloud's initial gravitational binding energy, so that the clouds would be bound if the thermal energy does not contribute significantly to the virial balance, and are therefore expected to form stars efficiently on their freefall timescale in the absence of feedback.\\
\section{Embedded cluster parameter space}
\begin{figure}
\includegraphics[width=0.5\textwidth]{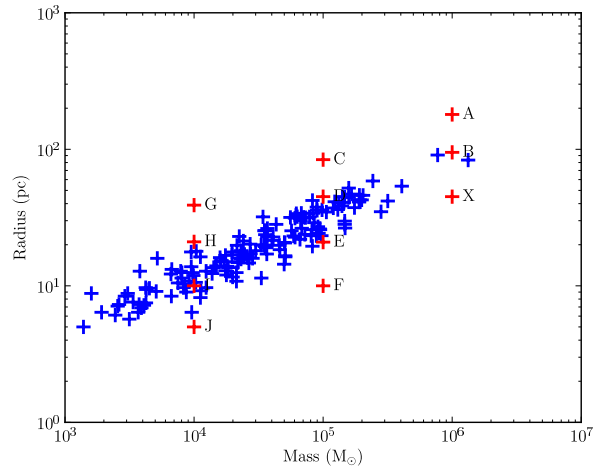}
\caption{Cluster mass--velocity dispersion parameter space with clouds from Heyer et al, 2009 plotted as blue crosses and our model clouds plotted as red crosses.} 
\label{fig:heyer_m_sigma}
\end{figure}

\begin{figure}
\includegraphics[width=0.5\textwidth]{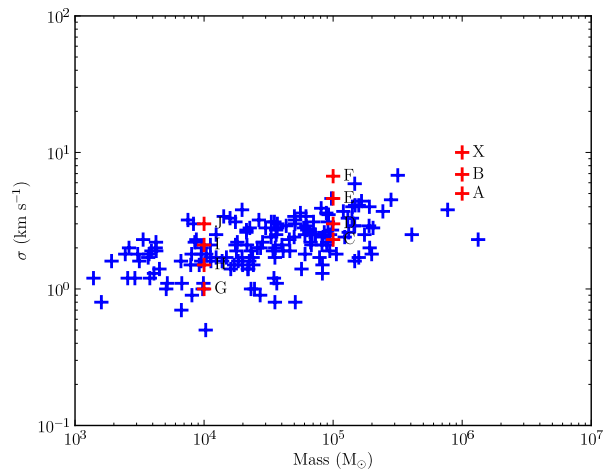}
\caption{Cluster mass--radius parameter space with clouds from Heyer et al, 2009 plotted as blue crosses and our model clouds plotted as red crosses.} 
\label{fig:heyer_m_r}
\end{figure}

\begin{figure}
\includegraphics[width=0.5\textwidth]{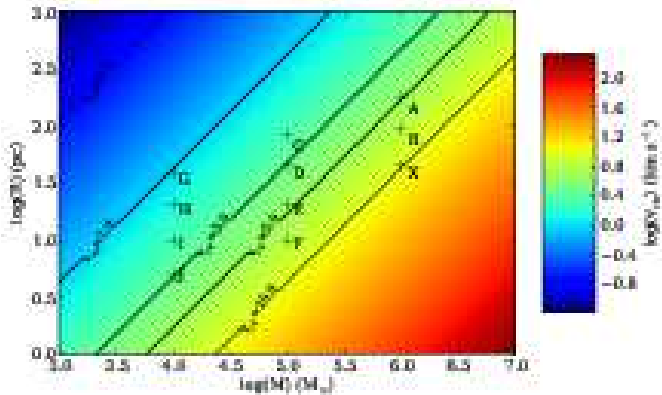}
\caption{Cluster mass--radius parameter space studied in this work. Colours and black contour lines are velocities required to give uniform clusters of given mass and radius a ratio $E_{\rm kin}/|E_{\rm pot}|$ of 0.7.} 
\label{fig:paramspace}
\end{figure}

\indent \cite{2009ApJ...699.1092H} derived masses, radii and velocity dispersions for 158 molecular clouds ranging in radius from a few to $\sim$100 pc and in mass from $\sim10^{3}$ to $\sim10^{6}$M$_{\odot}$. We chose the sizes and masses of our clouds to cover the higher mass end of \cite{2009ApJ...699.1092H}'s dataset, since GMC's with masses of $10^{3}$M$_{\odot}$ will not form many O--stars. We therefore study clouds in the mass range $10^{4}-10^{6}$ M$_{\odot}$, covering systems from approximately the size of Orion to 30--Doradus. We choose cloud radii in the range $5-$a few$\times10^{2}$pc, resulting in freefall times of between 0.8 and 20Myr. We choose initial turbulent velocities to in the range $1-10$km s$^{-1}$ so that all of our clouds have the same ratio $E_{\rm kin}/|E_{\rm pot}|$ of 0.7. In Figures \ref{fig:heyer_m_sigma} and \ref{fig:heyer_m_r} mass--radius and mass--velocity dispersion plots with our simulated clusters overlaid on the data from \cite{2009ApJ...699.1092H}, showing that our clouds overlap nicely the parameters of observed star--forming clouds. In Figure \ref{fig:paramspace}, we plot the mass--radius parameter space with colours and black contour lines overlaid representing the velocity required to give each cluster our chosen initial virial ratio.\\
\indent This picture is complicated somewhat by our equation of state. If we were to make the simple canonical assumption that the gas in our clouds is purely isothermal with an average temperature of 10K, the thermal energy would make a negligible contribution to the clouds' energy balance in all cases. This is not necessarily true when we make use of the more realistic Larson equation of state, since the gas densities are initially low enough that the gas is in the warm non--isothermal regime, so that the contribution of the thermal energy may become important.\\
\indent In Table \ref{tab:init} we give the total mass, initial radius, initial RMS turbulent velocity, initial mean number density (assumed to be molecular), freefall time, mean initial temperature and true virial ratio (including the thermal energy) of all of our clouds (mean quantities are mass--weighted). The initial mean temperatures of several of the clouds are very high and, in the case of Runs G and H, the initial thermal energy of the cloud is large enough that the clouds are not initially bound. In the cases of clouds A, C and G, the initial mean temperatures are sufficiently high that the clouds should more properly be regarded as atomic and not molecular. However, all these clouds are seeded with supersonic turbulent velocity fields which will shock the gas, locally increasing the density and decreasing the temperature. It is therefore not obvious simply from the contents of Table \ref{tab:init} that even the warmest clouds will not form at least some stars.\\
\begin{table*}
\begin{tabular}{|l|l|l|l|l|l||l|l|l|}
Run&Mass (M$_{\odot}$)&Radius (pc)&v$_{\rm RMS}$ (km s$^{-1}$)&$\langle$ n(H$_{2}$) $\rangle$ (cm$^{-3}$)&t$_{\rm ff}$ (Myr)&$\langle$ T(K) $\rangle$&(E$_{\rm kin}$+E$_{\rm therm}$)/$|$E$_{\rm pot}|$\\
\hline
A&$10^{6}$&180&5.0&2.9&19.6&143&0.72\\
\hline
B&$10^{6}$&95&6.9&16&7.50&91&0.70\\
\hline
X&$10^{6}$&45&9.6&149&2.44&52&0.69\\
\hline
\hline
C&$10^{5}$&84&2.3&2.6&19.7&145&0.90\\
\hline
D&$10^{5}$&45&3.0&15&7.70&92&0.78\\
\hline
E&$10^{5}$&21&4.6&147&2.46&52&0.71\\
\hline
F&$10^{5}$&10&6.7&1439&0.81&30&0.69\\
\hline
\hline
G&$10^{4}$&39&1.0&2.2&19.7&149&1.92\\
\hline
H&$10^{4}$&21&1.5&14&7.79&93&1.10\\
\hline
I&$10^{4}$&10&2.1&136&2.56&53&0.79\\
\hline
J&$10^{4}$&5&3.0&1135&0.90&32&0.72\\
\end{tabular}
\caption{Initial properties (mass, radius, turbulent velocity dispersion, mean initial molecular number density, freefall time, mean initial temperature and virial ratio) of all runs.}
\label{tab:init}
\end{table*}
\indent For those clouds that are able to form stars, we model the effect of ionization feedback. Photoionization and winds act for the whole duration of a massive star's lifetime, whereas supernovae are isolated events occurring at the end of that lifetime. The influence of feedback is likely to be dominated by the most massive stars, which have main--sequence lifetimes of $\sim3$Myr. This sets the timescale on which photoionization and winds can modify GMCs before the action of supernovae. In about half the clouds studied here -- runs X, E, F, I and J -- the cloud freefall time is less than or comparable to 3Myr, which we will refer to in future as $t_{\rm SN}$. If star formation is to go to completion and these clouds be disrupted in a few crossing times, most or all of the work must be done by photoionization and winds/jets (note that we defer the study of the effects of a winds to a later paper).\\
\section{Regulation of photoionizing feedback}
\begin{figure}
\includegraphics[width=0.5\textwidth]{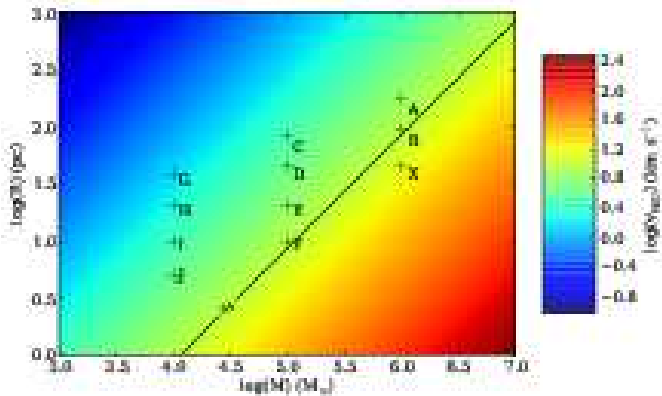}
\caption{Escape velocities of clusters studied in this work.} 
\label{fig:vesc}
\end{figure}

\indent There are at first sight three issues which determine how strong the effects of ionizing feedback may be on a given cluster:\\
\indent (i) Number and luminosity of ionizing sources. The number of ionizing sources present at a given time in a cluster's evolution depends on the mass function and the total stellar mass. The dominant sources of ionizing feedback are the most massive stars or clusters and, since the stellar and cluster mass functions are steep power laws, their high--mass ends are likely to be poorly sampled and subject to stochastic effects whereby a small change in the initial conditions or input physics may result in the same quantity of mass being distributed amongst different numbers and (therefore masses) of individual objects. Since the ionizing luminosity of stars and clusters are rather strong functions of their masses, this may in principle lead to statistical uncertainties in the total ionizing flux and how the sources are distributed at a given time in a simulation. In an attempt to eliminate some of the stocahsticity inherent in star formation we do not enable ionization by clusters in our $10^{5}$ and $10^{6}$M$_{\odot}$ calculations until respectively three and ten clusters hosting at least one O--star have formed, and in the $10^{4}$M$_{\odot}$ runs until three ionizing sources have formed. We subsequently find, as detailed in the Appendix, that allowing ionization to begin as soon as any ionizing sources have formed actually does not strongly affect the results. The reason for this is largely that variations of a factor of a few in the ionizing luminosity at any given time do not strongly affect the evolution of our clouds, as also shown explicitly in the Appendix.\\
\indent (ii) The ability of sources to ionize fresh gas. The rate at which new material is ionized is expected to depend strongly on the ambient gas density and on the strength of accretion flows which may swamp ionizing sources with neutral material \citep[e.g][]{2010ApJ...719..831P,2011MNRAS.414..321D}. \cite{1995RMxAC...1..137W} gives a simple formula for a critical accretion rate $\dot{M}_{\rm crit}$ onto an ionizing source of mass $M$ and ionizing luminosity $Q_{\rm H}$ (Lyman continuum photons per second) above which all the ionizing photons are absorbed by the accretion flow, so that the HII region cannot expand and may instead contract:
\begin{eqnarray}
\dot{M}>\left(\frac{4\pi Q_{\rm H}GMm^{2}_{\rm H}}{\alpha_{\rm B}}\right)^{\frac{1}{2}}.
\label{eqn:mdotcrit}
\end{eqnarray}
The accretion rate onto a given source depends on the ambient gas density and may be roughly estimated from the Bondi accretion rate $\dot{M}_{\rm B}$, given by
\begin{eqnarray}
\dot{M}_{\rm B}=\frac{4\pi G^{2}M^{2}\rho}{c_{\rm s}^{3}}
\end{eqnarray}
If we crudely set $\dot{M}_{\rm crit}=\dot{M}_{\rm B}$ and set $c_{\rm s}=c_{\rm II}$ (the speed of sound in ionized gas), we obtain the condition that, for an accretion flow to swamp an ionizing source
\begin{eqnarray}
\frac{GM}{c_{\rm II}^{2}}>\left(\frac{Q_{\rm H}}{4\pi n^{2}\alpha_{\rm B}}\right)^{\frac{1}{3}},
\label{eqn:resc}
\end{eqnarray}
which is, to within a factor close to unity, the same as saying that the radius at which the escape velocity exceeds the sound speed in the ionized gas must exceed the Str\"omgren radius in order for the HII region to be trapped. (Note that we do not artificially shut off sources whose accretion rates exceed these critical values -- all ionizing sources are left to contend with their accretion flows self--consistently.)\\
\indent (iii) The ability of ionized gas to expel neutral material. Unless a cluster's O--stars are able to ionize its entire reserve of neutral gas, the ability of the massive stars to disrupt the system will depend on how effectively the expanding HII regions can sweep up neutral material. This in turn depends largely on the escape velocity of the system as a whole (as opposed to the escape velocity of individual stars or subclusters, which governs how much gas is likely to be ionized) compared to the sound speed in ionized gas. In Figure \ref{fig:vesc}, we plot the variation in escape velocity across our parameter space, with a contour indicating $c_{\rm II}$ which we take to be 10 km s$^{-1}$. Evidently, some of the denser clusters in the parameter space have such high escape velocities that ionization will struggle to expel gas from them.
\captionsetup[subfigure]{labelformat=empty}
\begin{figure*}
     \centering
     \subfloat[Run A]{\includegraphics[width=0.30\textwidth]{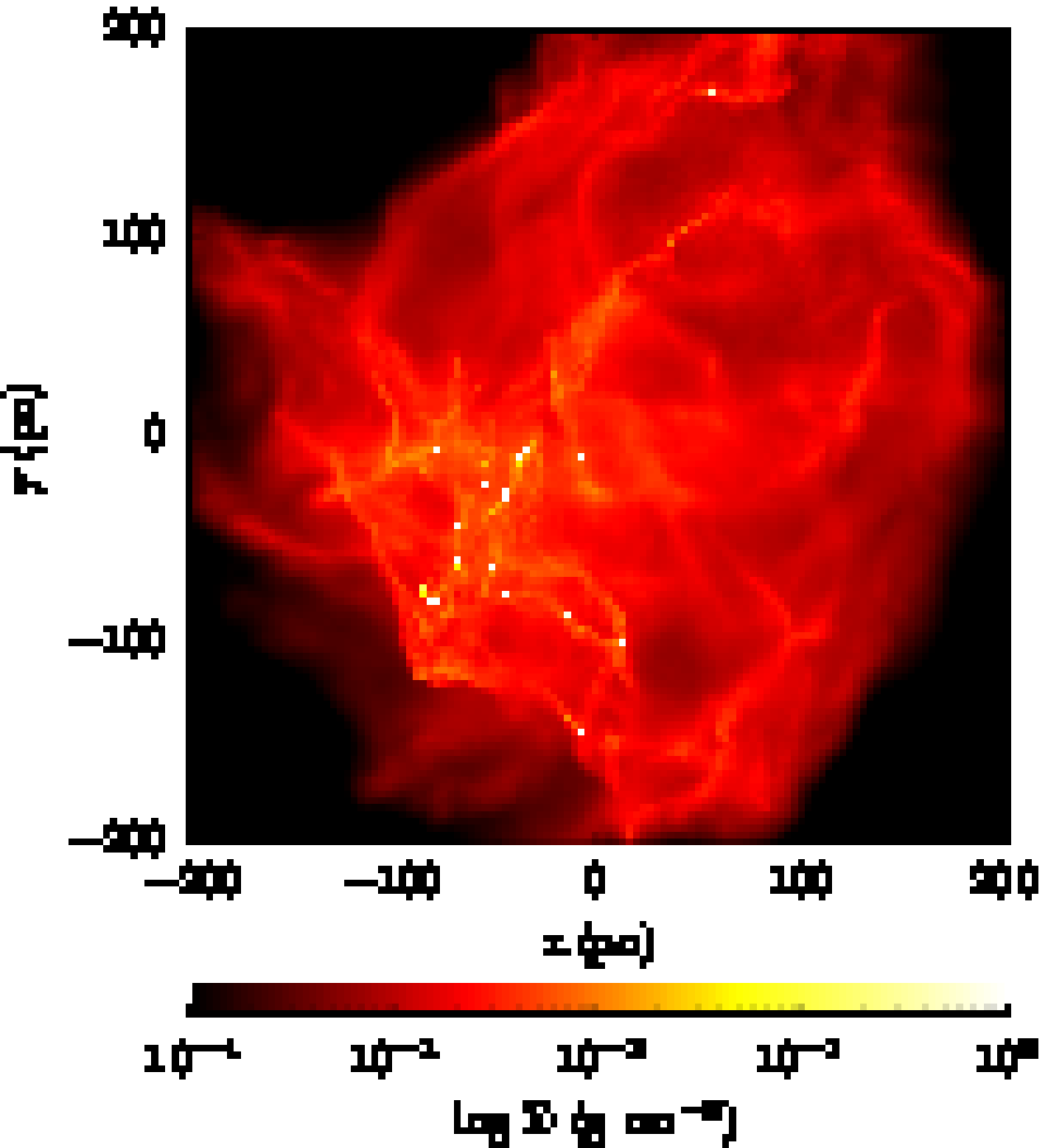}}     
     \hspace{.1in}
     \subfloat[Run B]{\includegraphics[width=0.30\textwidth]{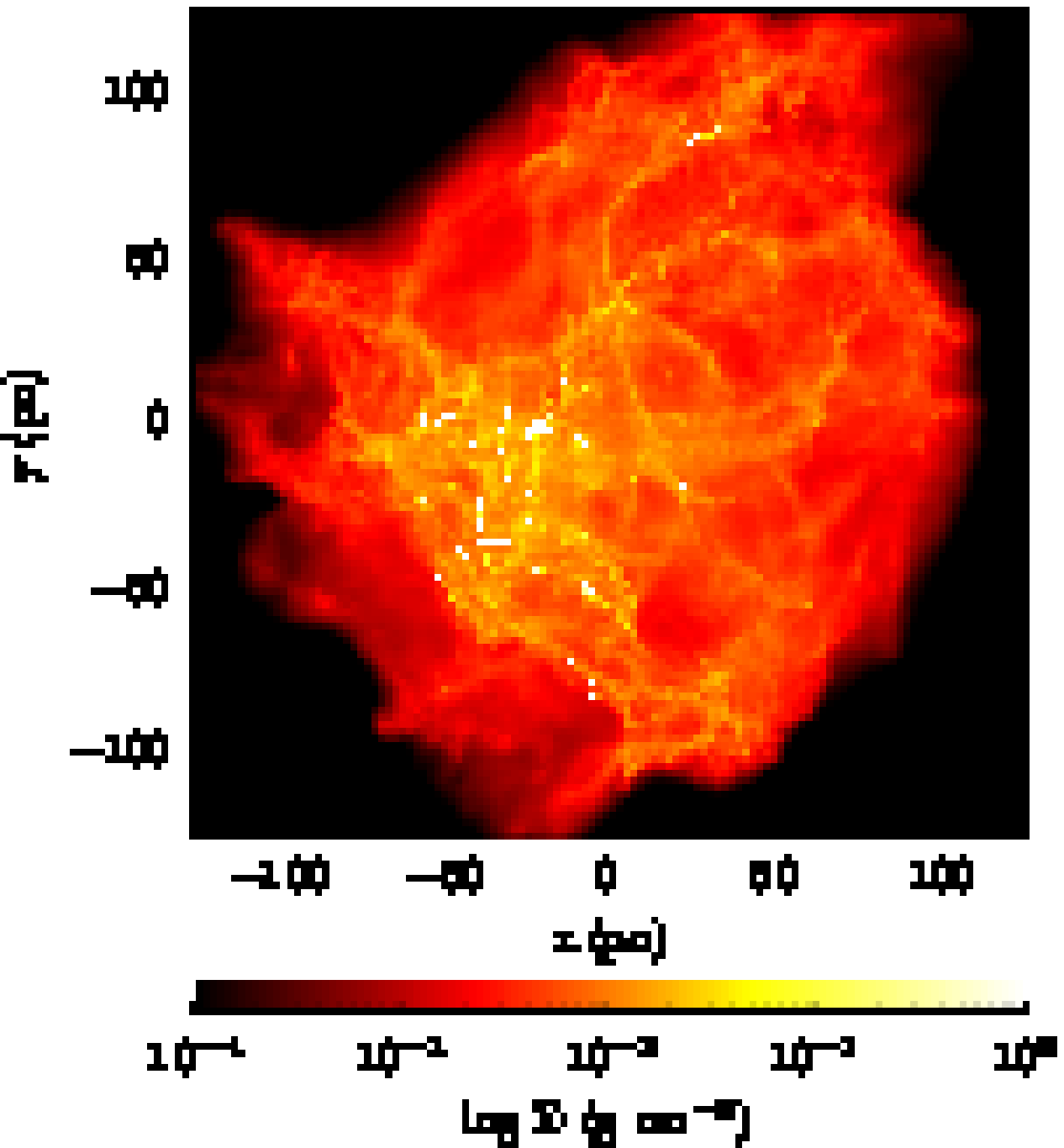}}
     \hspace{.1in}
     \subfloat[Run X]{\includegraphics[width=0.30\textwidth]{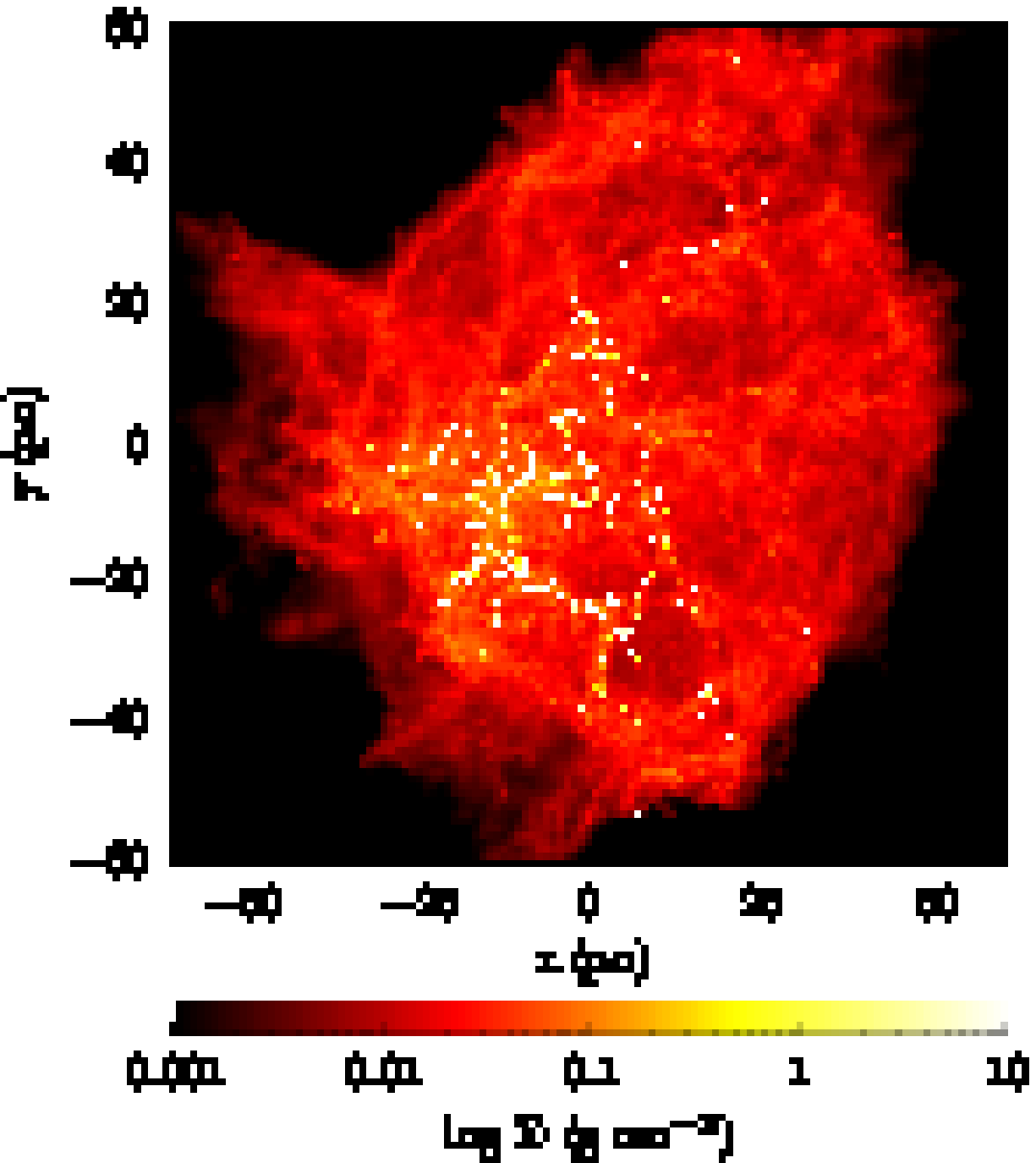}}
          \vspace{.1in}
     \subfloat[Run D]{\includegraphics[width=0.30\textwidth]{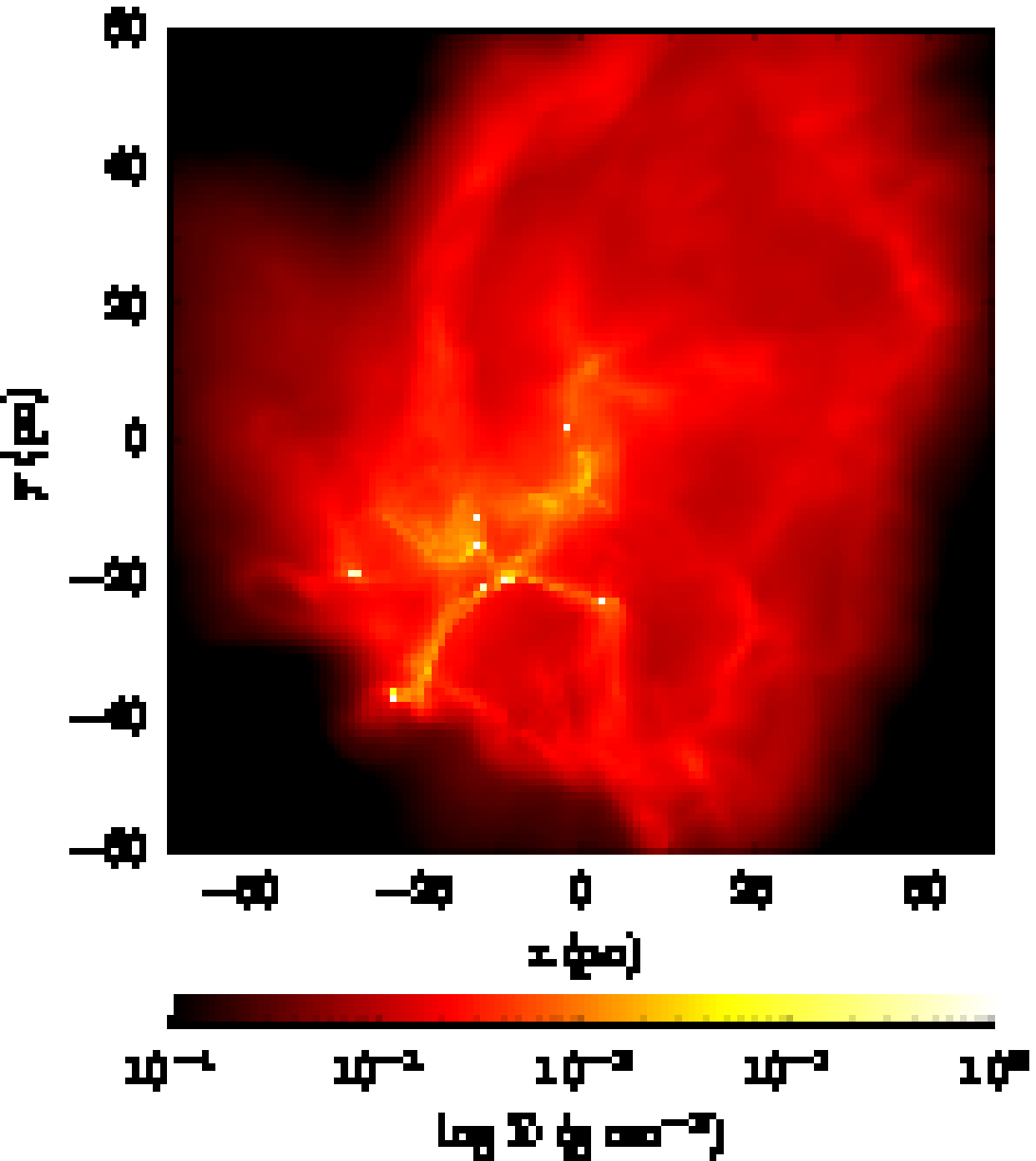}}
     \hspace{.1in}
     \subfloat[Run E]{\includegraphics[width=0.30\textwidth]{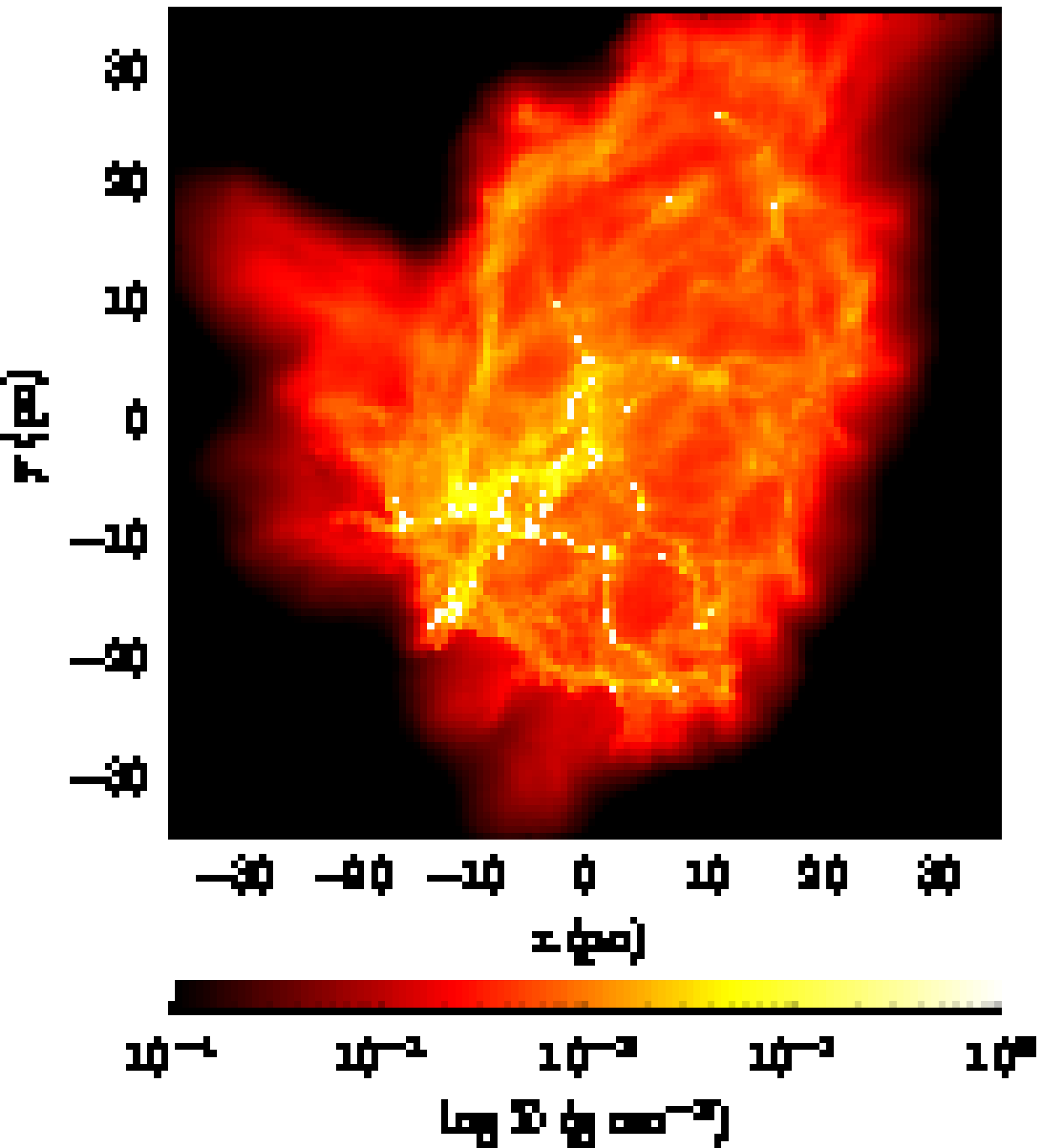}}
          \hspace{.1in}
     \subfloat[Run F]{\includegraphics[width=0.30\textwidth]{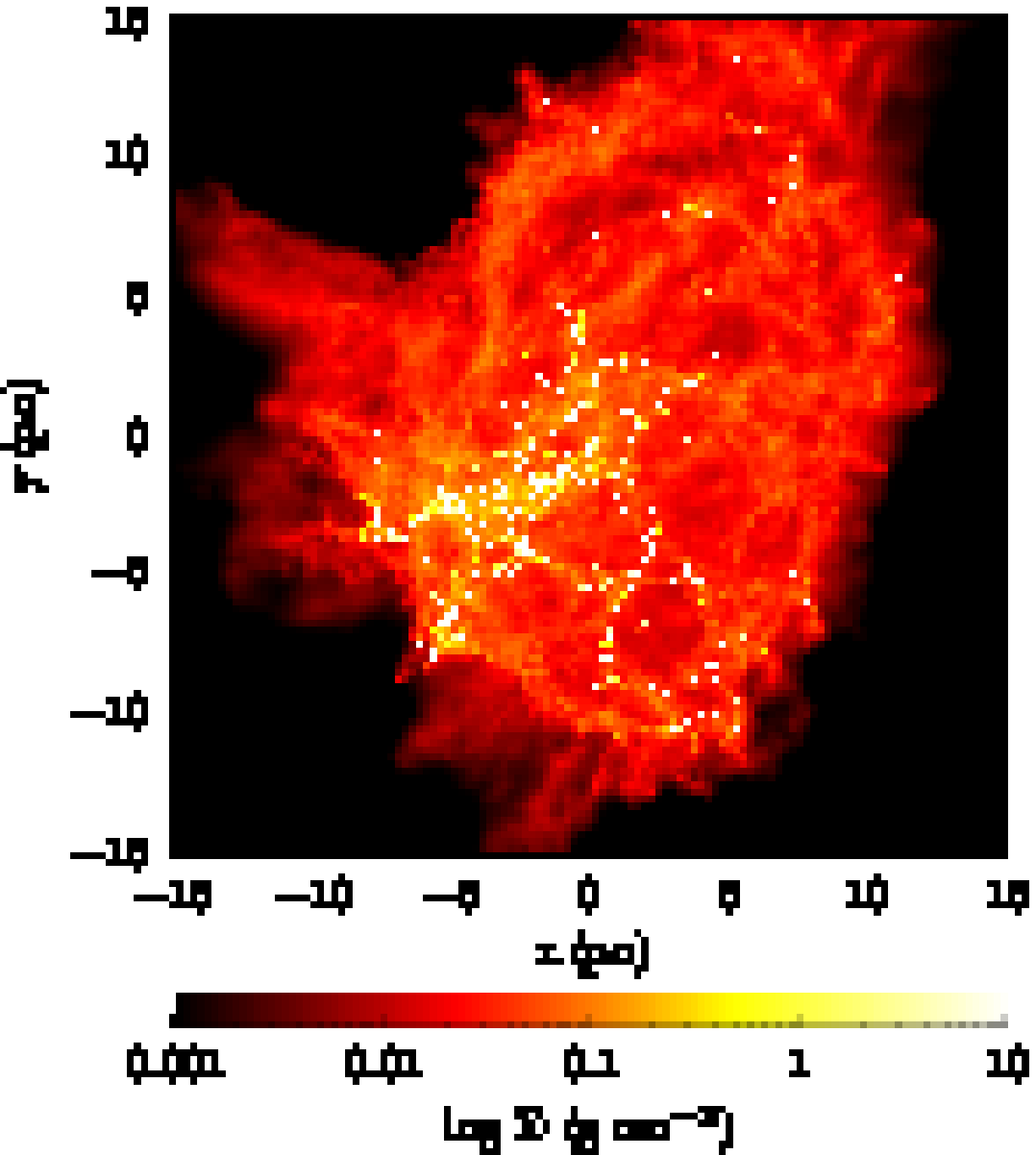}}
          \vspace{.1in}
     \subfloat[Run I]{\includegraphics[width=0.30\textwidth]{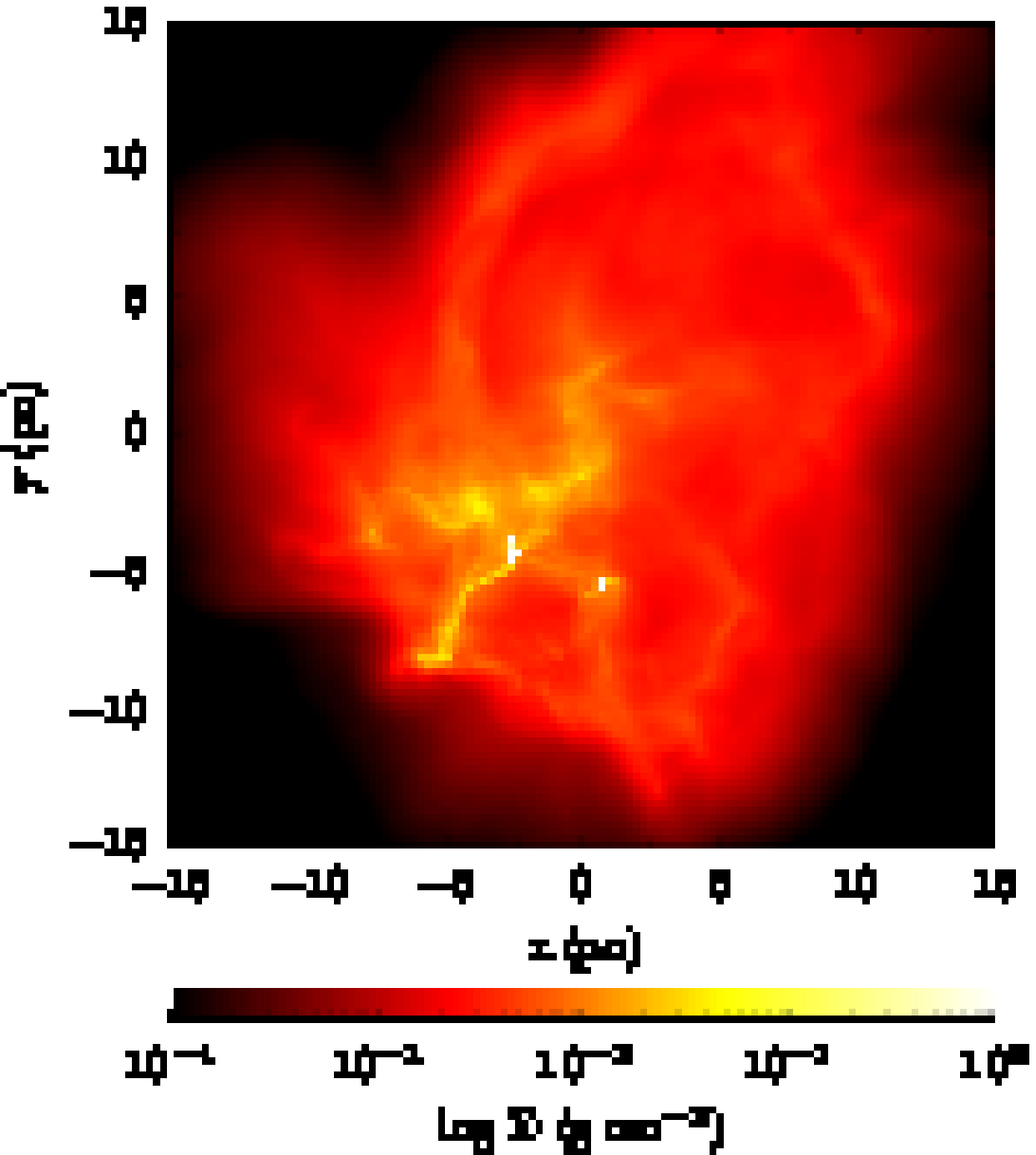}}
          \hspace{.1in}
     \subfloat[Run J]{\includegraphics[width=0.30\textwidth]{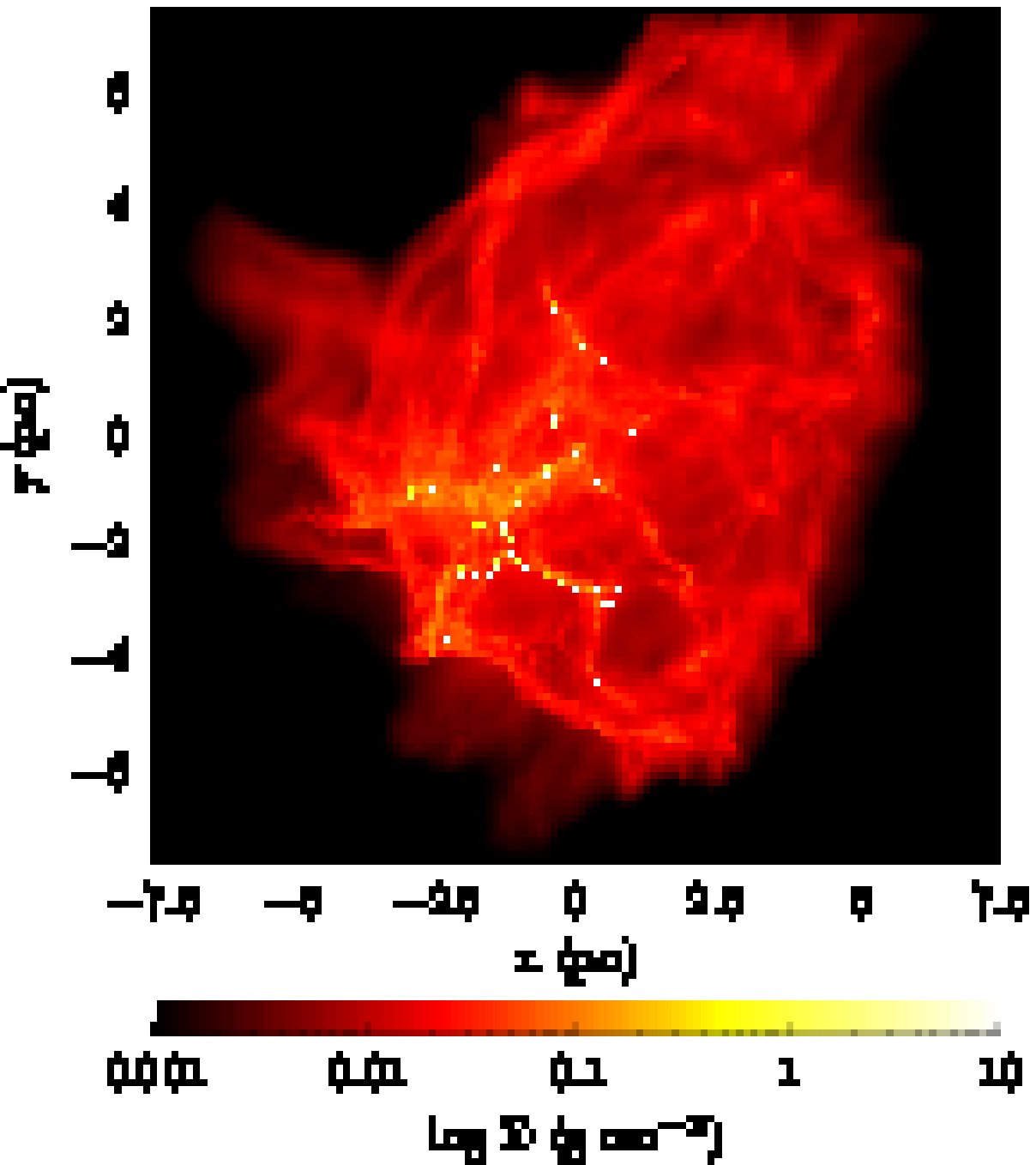}}
     \caption{Gallery of initial conditions of clusters, as shown by column density maps observed down the $z$--axis. White dots represent sink particles (individual stars in Runs I and J, clusters otherwise) and are not to scale. Note the different physical sizes and the different column density scales.}
   \label{fig:gallery}
\end{figure*}   
\begin{figure*}
     \centering
     \subfloat[Run A]{\includegraphics[width=0.30\textwidth]{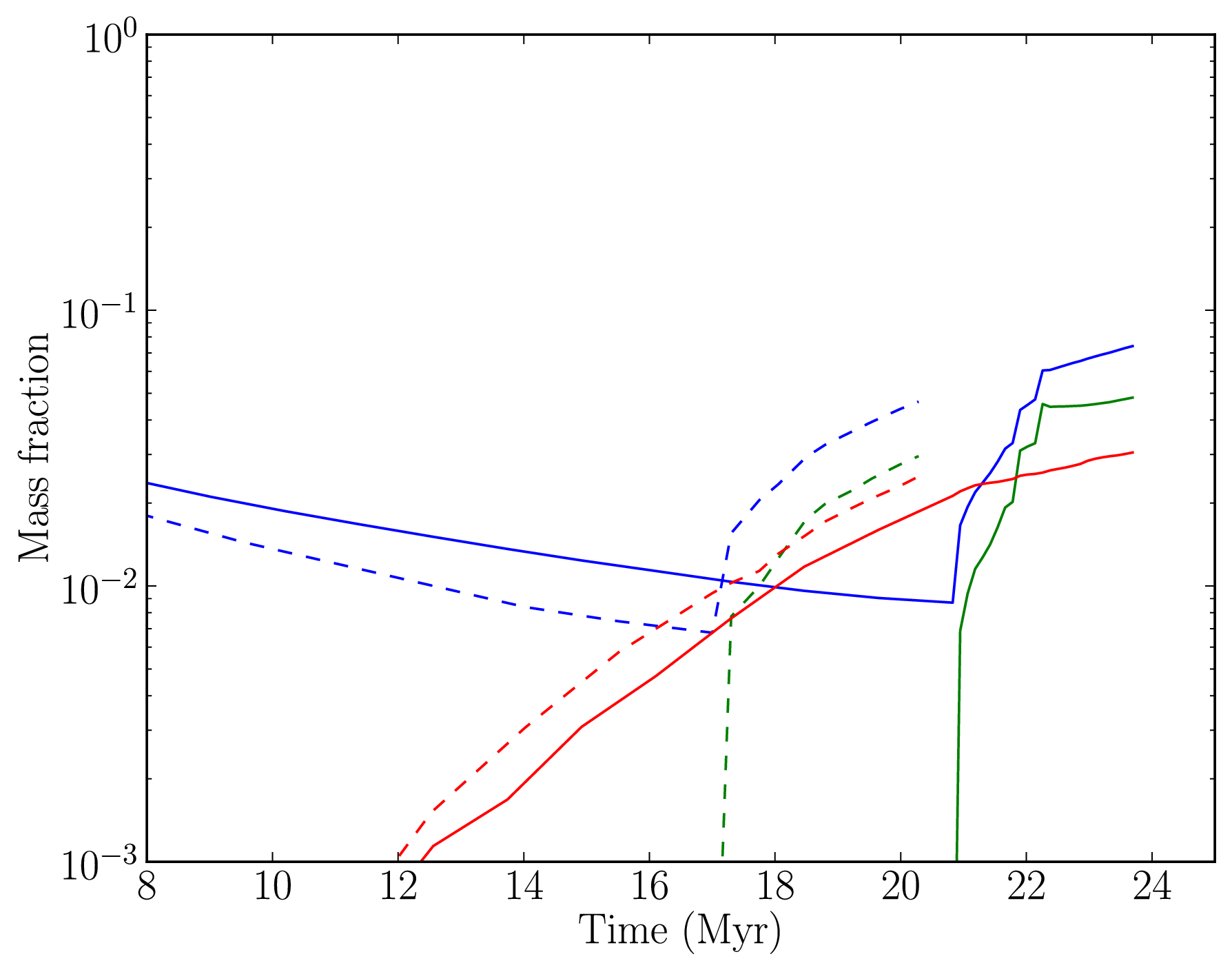}}     
     \hspace{.1in}
     \subfloat[Run B]{\includegraphics[width=0.30\textwidth]{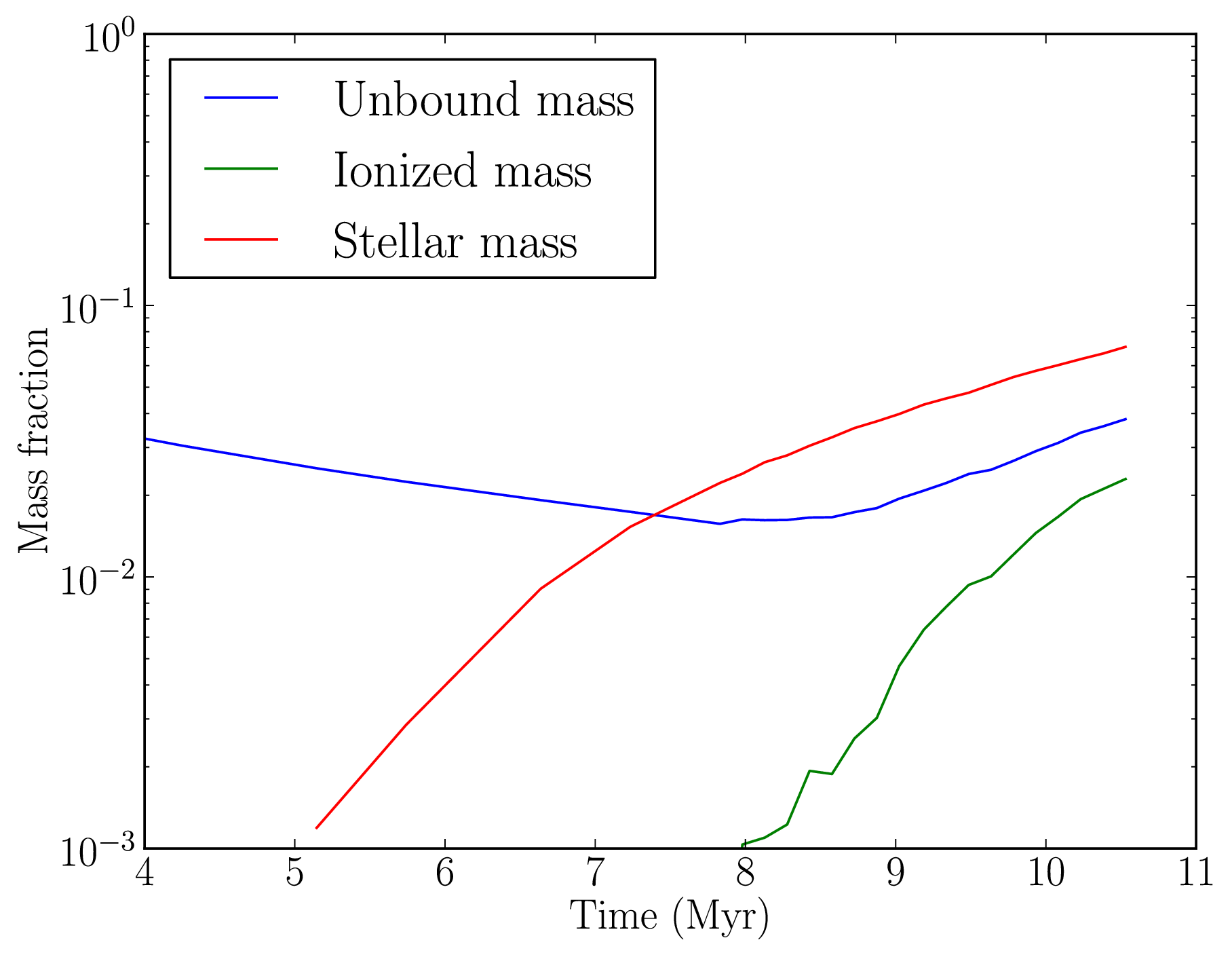}}
     \hspace{.1in}
     \subfloat[Run X]{\includegraphics[width=0.30\textwidth]{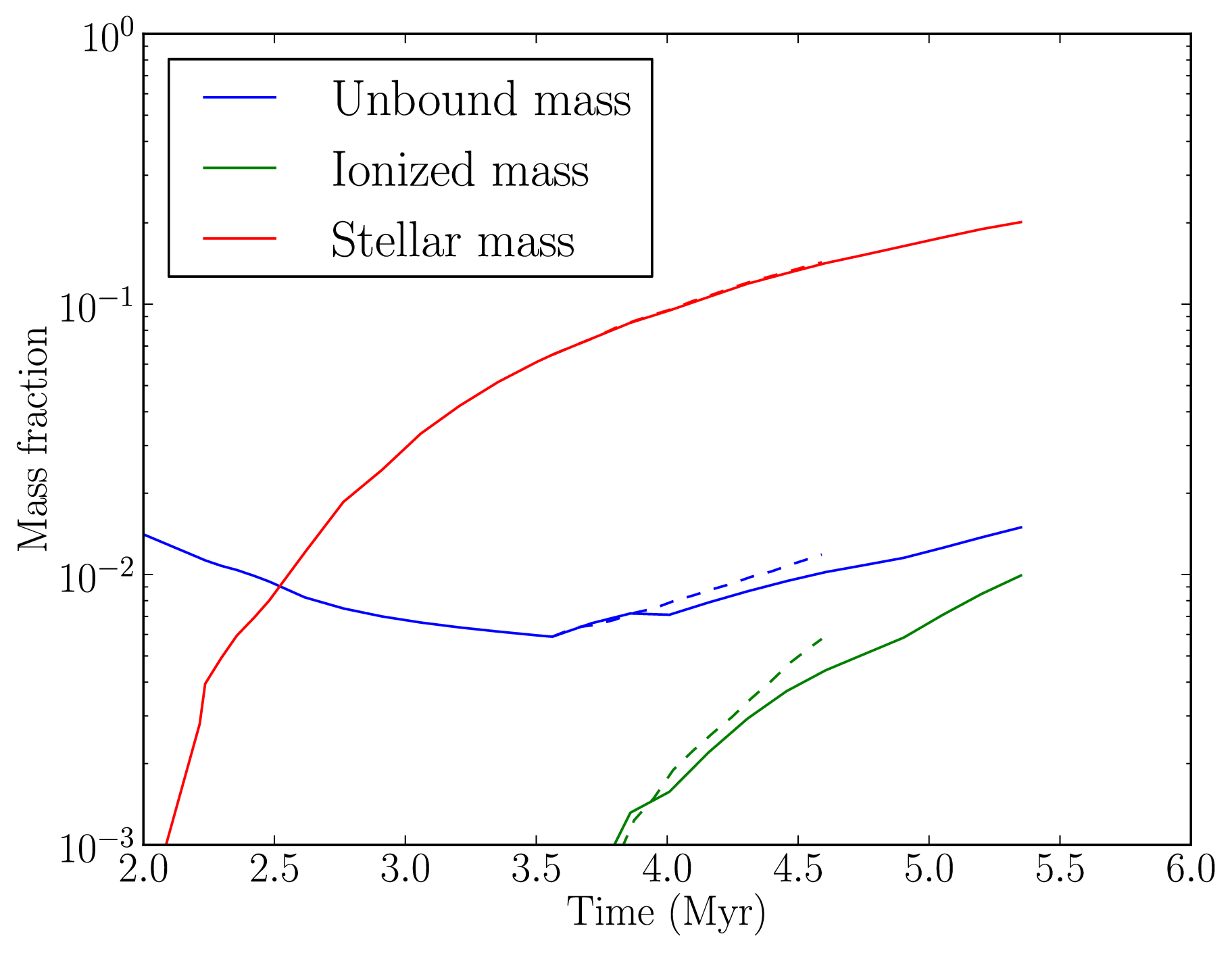}}
          \vspace{.1in}
     \subfloat[Run D]{\includegraphics[width=0.30\textwidth]{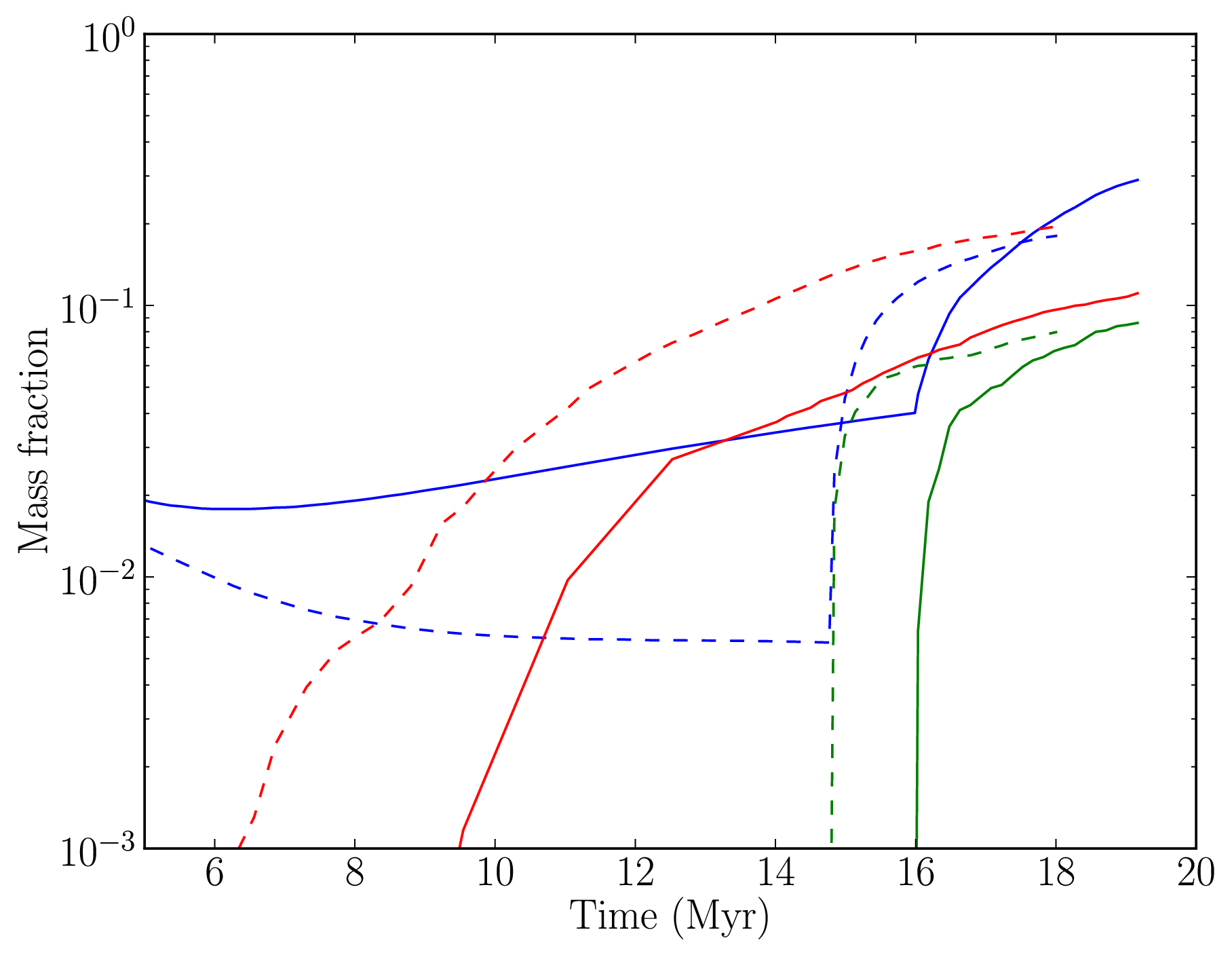}}
     \hspace{.1in}
     \subfloat[Run E]{\includegraphics[width=0.30\textwidth]{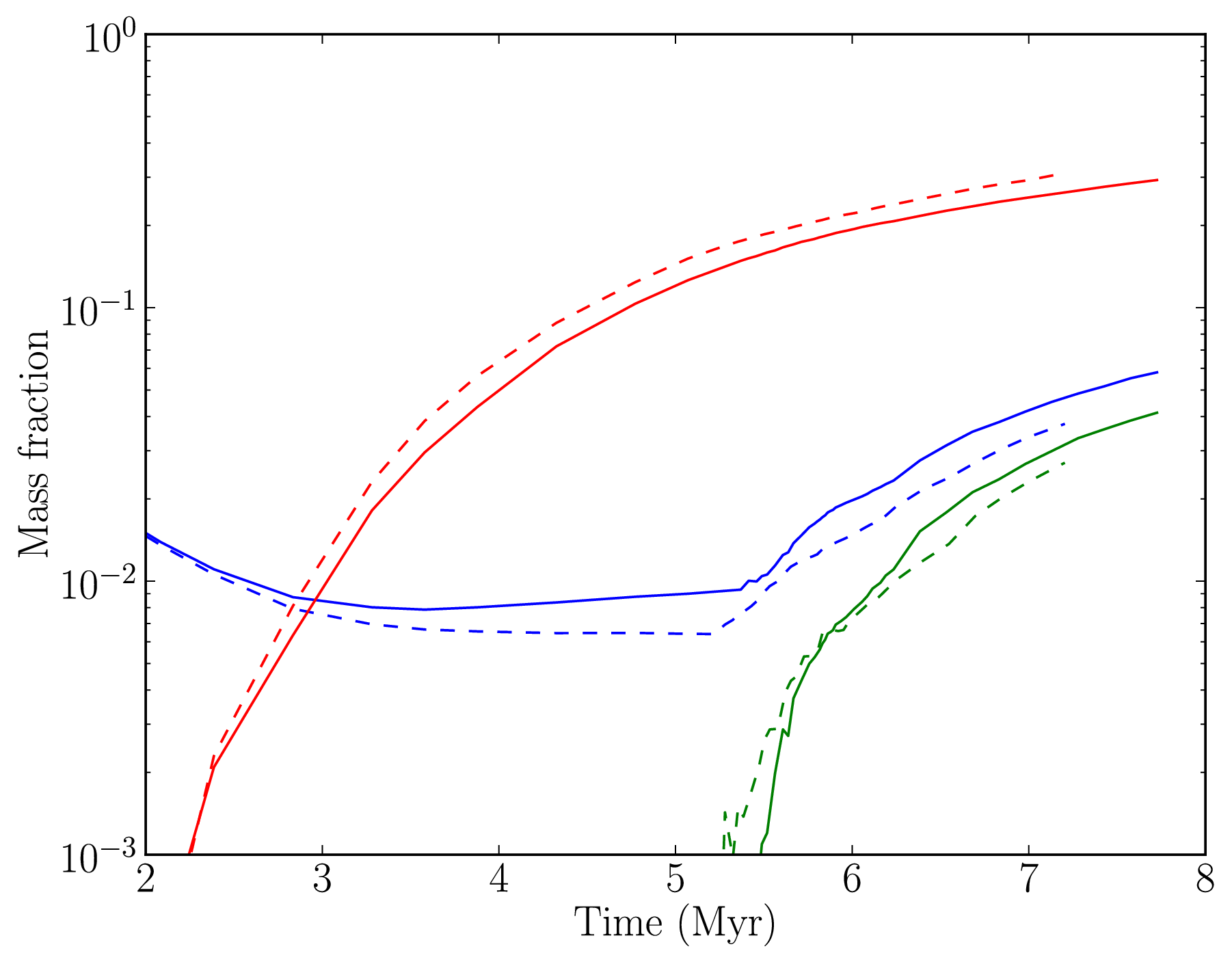}}
          \hspace{.1in}
     \subfloat[Run F]{\includegraphics[width=0.30\textwidth]{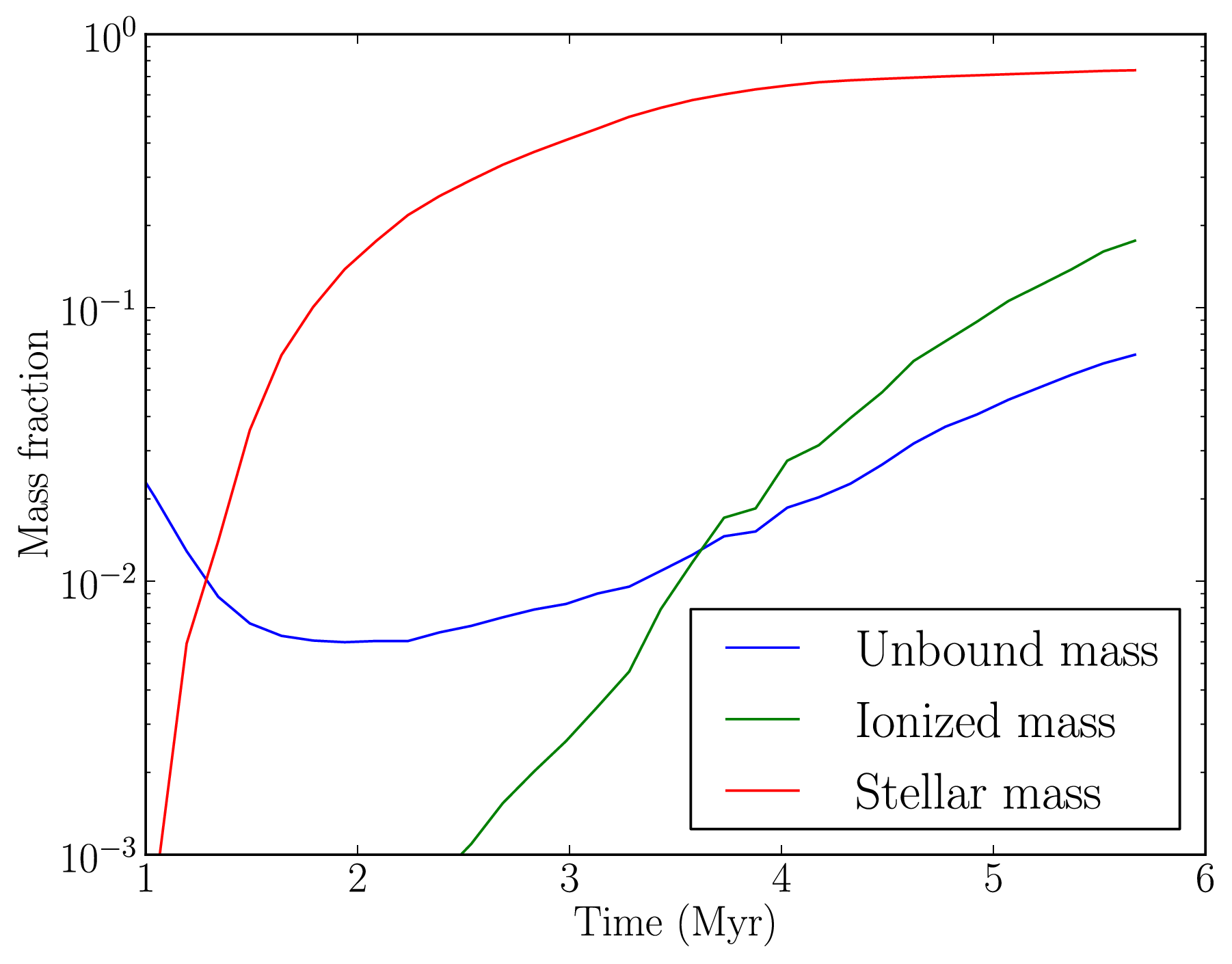}}
          \vspace{.1in}
     \subfloat[Run I]{\includegraphics[width=0.30\textwidth]{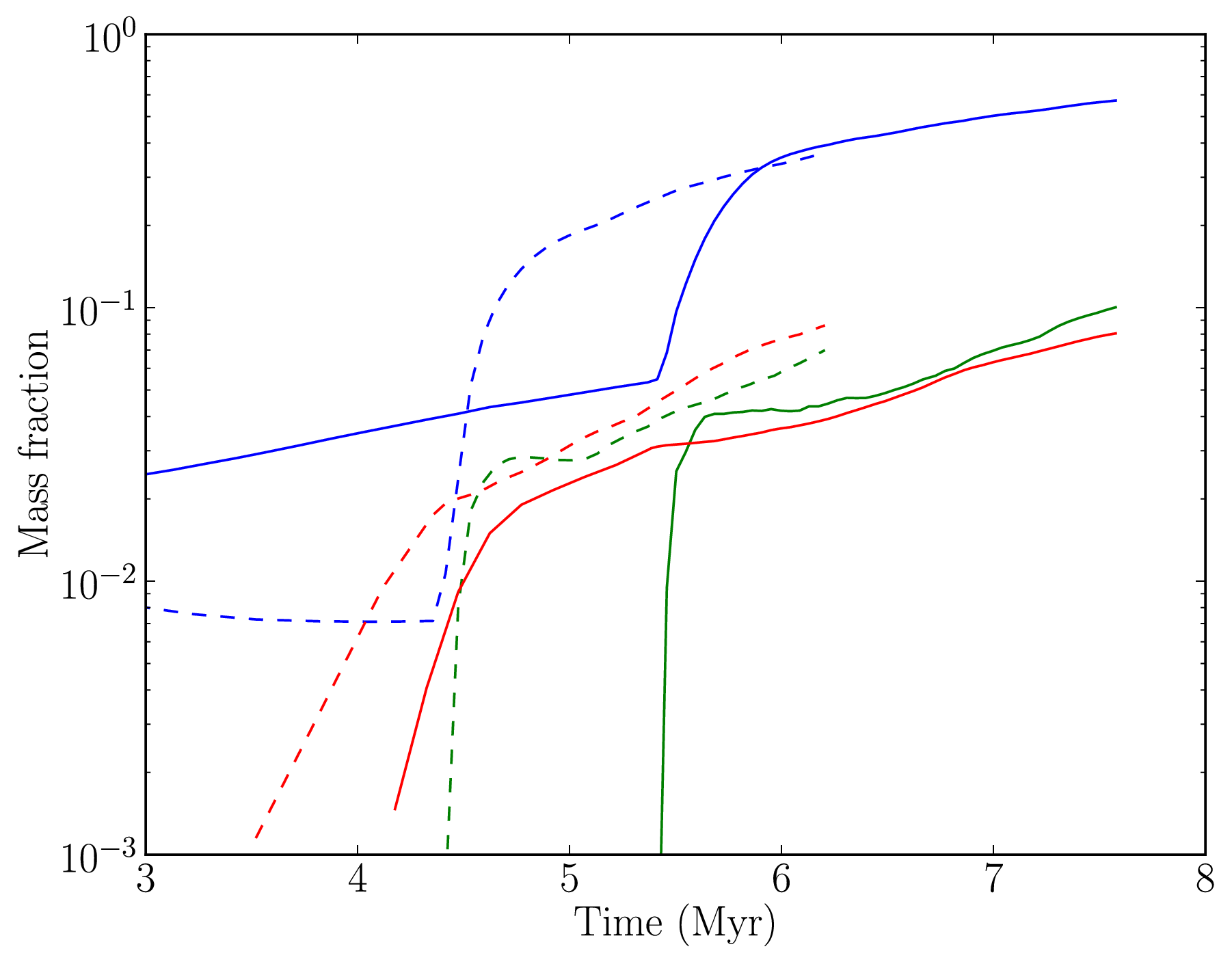}}
          \hspace{.1in}
     \subfloat[Run J]{\includegraphics[width=0.30\textwidth]{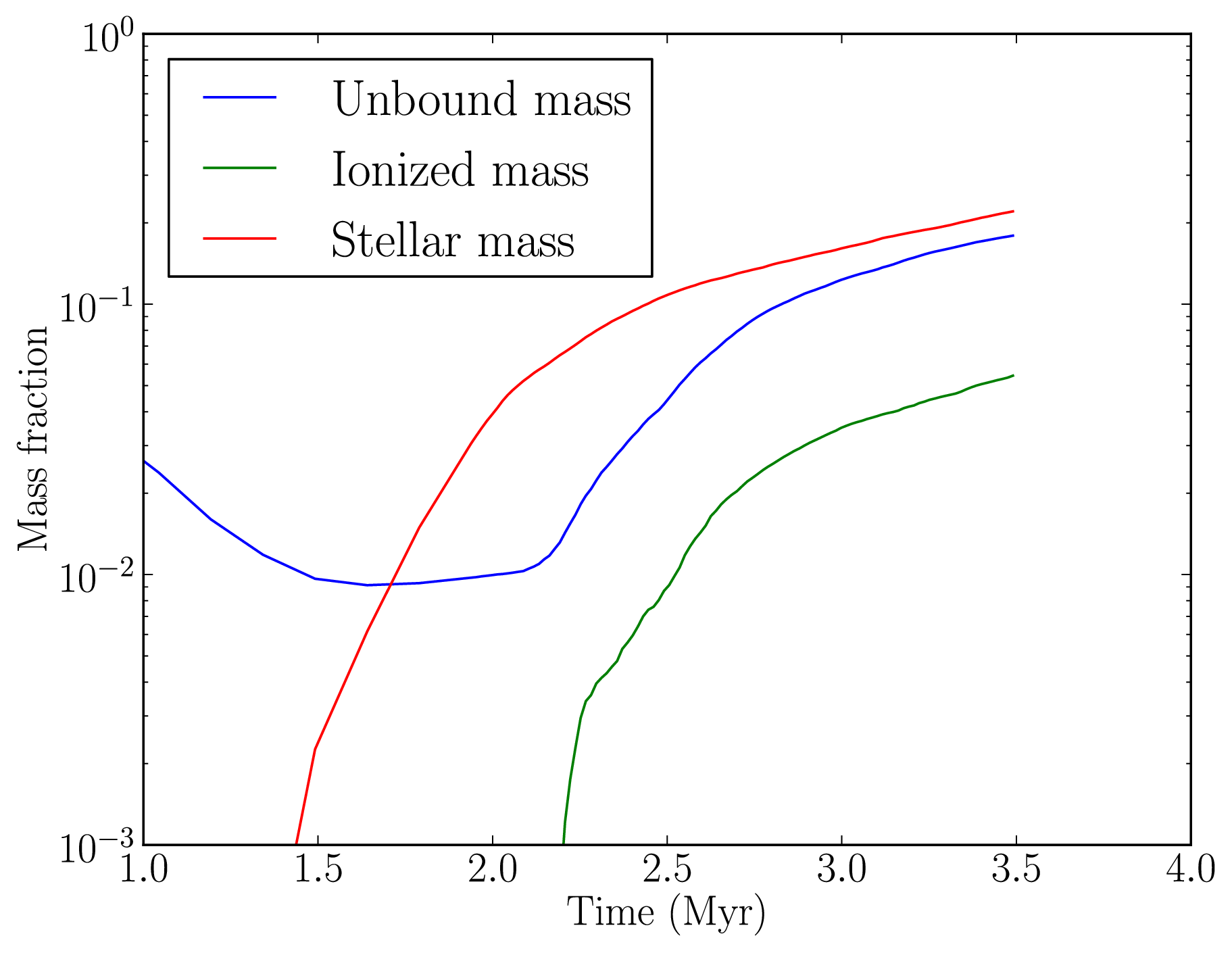}}
     \caption{Plots showing the evolution with time of the star formation efficiency (red), ionized gas fraction (green) and ubound mass fraction (blue) in all simulations. Solid lines are for standard runs using the Larson equation of state. Dashed lines in Runs A, D, E and I are for runs using an isothermal equation of state with a temperature of 10K. Dashed lines in Run X are from a repeat of part of the simulation from Dale \& Bonnell, 2011, in which our older, less accurate multiple--source ionization code was used. Note the different horizontal scaling on the plots.}
   \label{fig:unbnd}
\end{figure*}  
\begin{figure*}
     \centering
     \subfloat[Run A]{\includegraphics[width=0.30\textwidth]{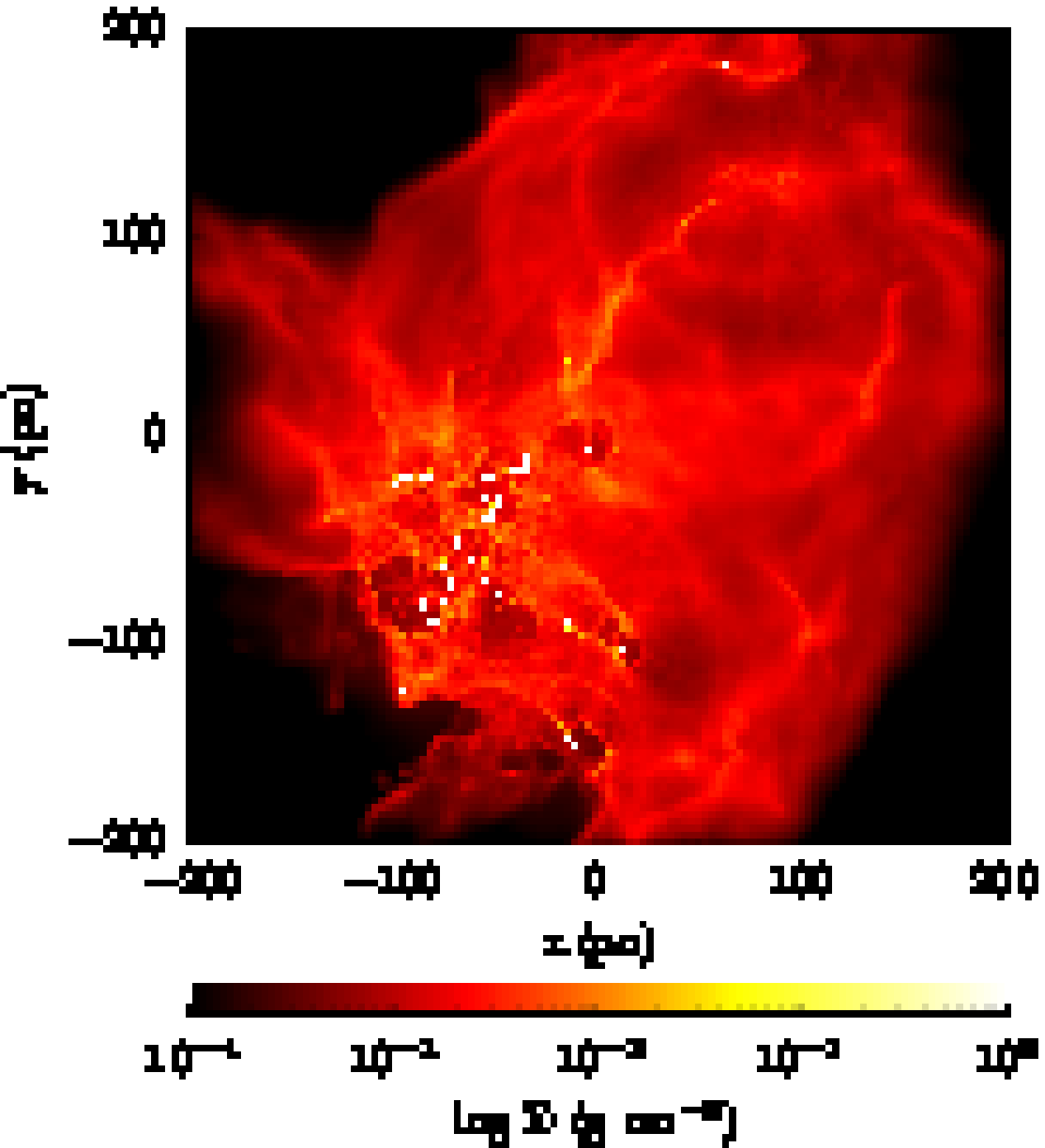}}     
     \hspace{.1in}
     \subfloat[Run B]{\includegraphics[width=0.30\textwidth]{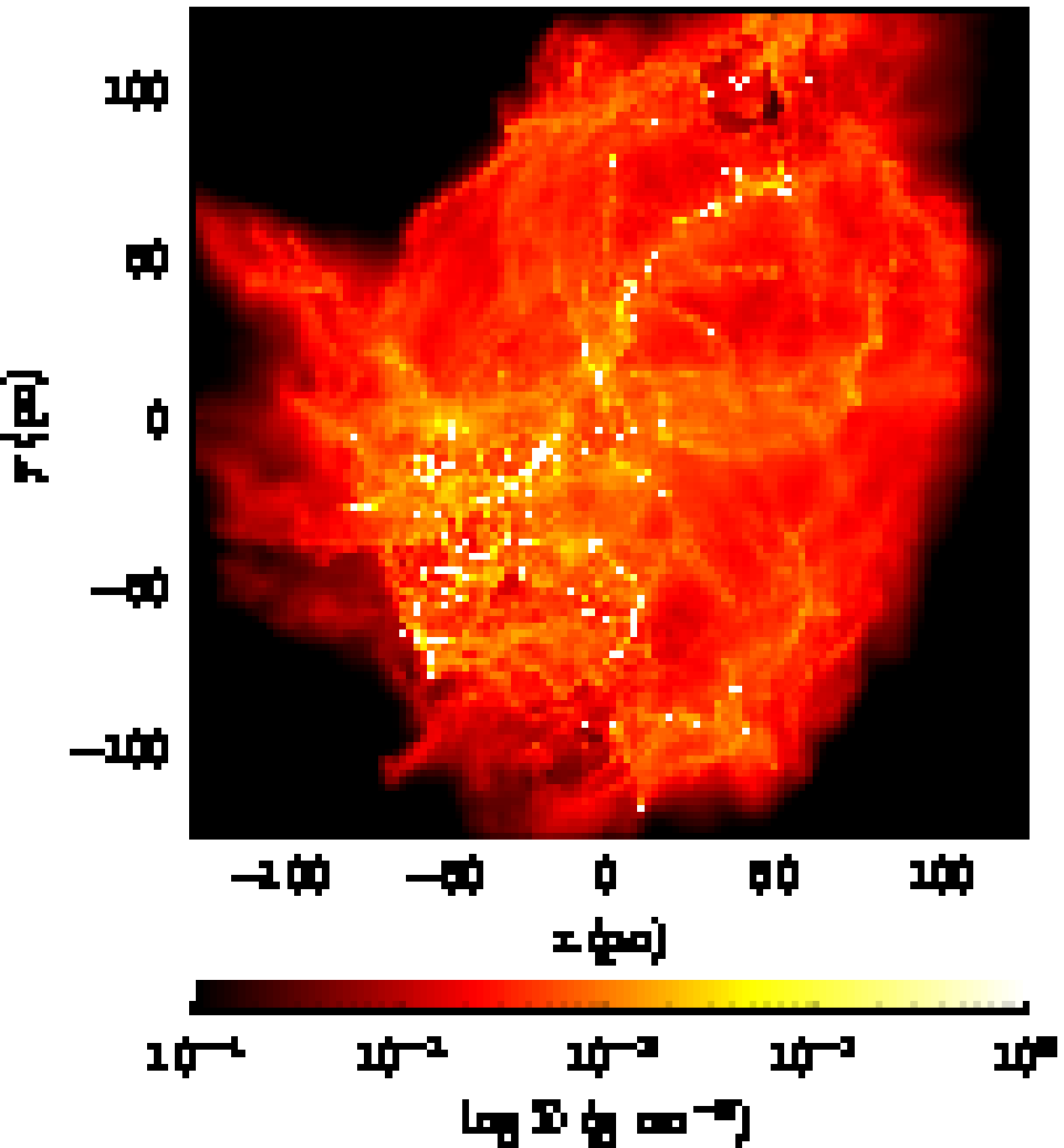}}
     \hspace{.1in}
     \subfloat[Run X]{\includegraphics[width=0.30\textwidth]{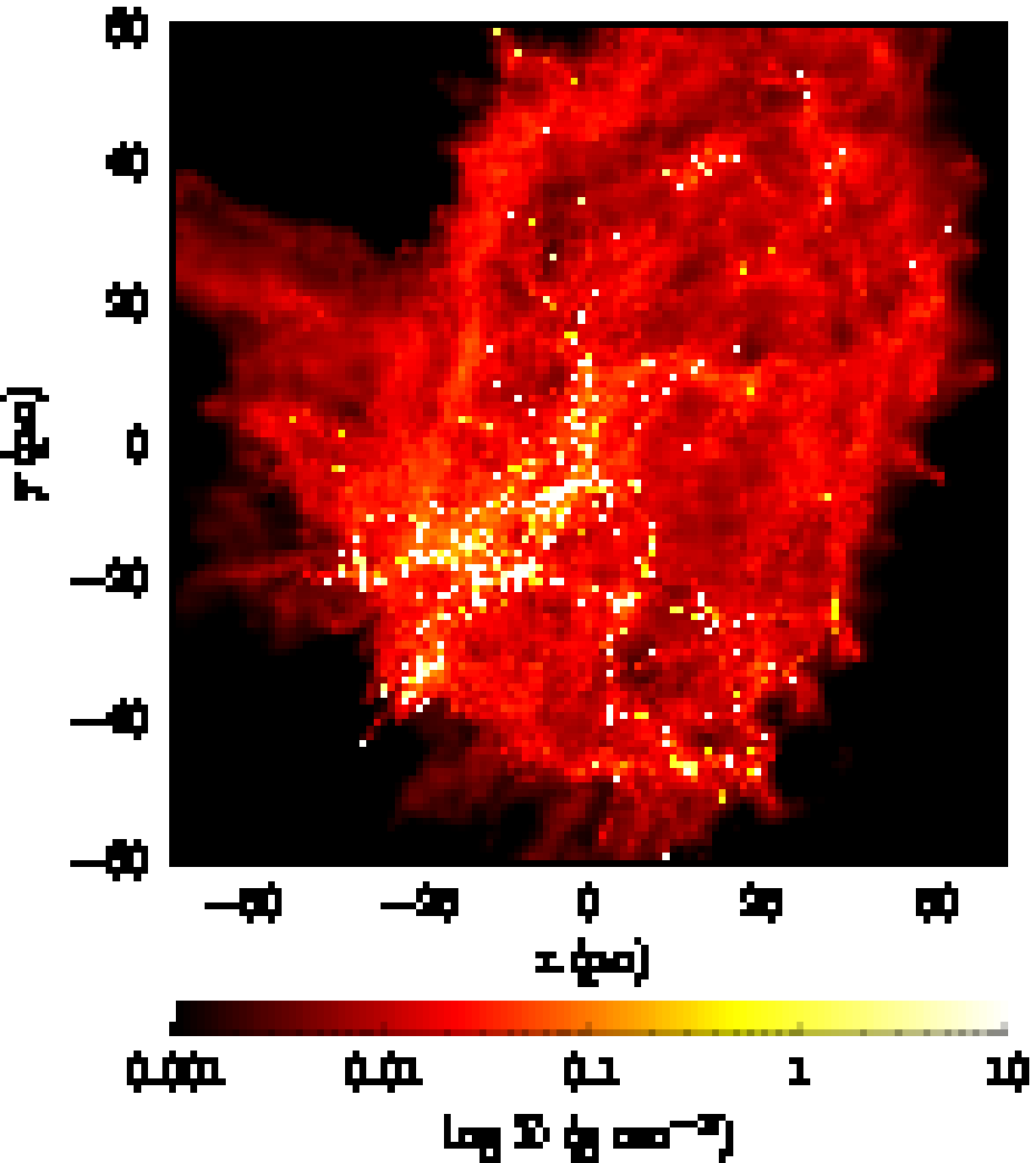}}
          \vspace{.1in}
     \subfloat[Run D]{\includegraphics[width=0.30\textwidth]{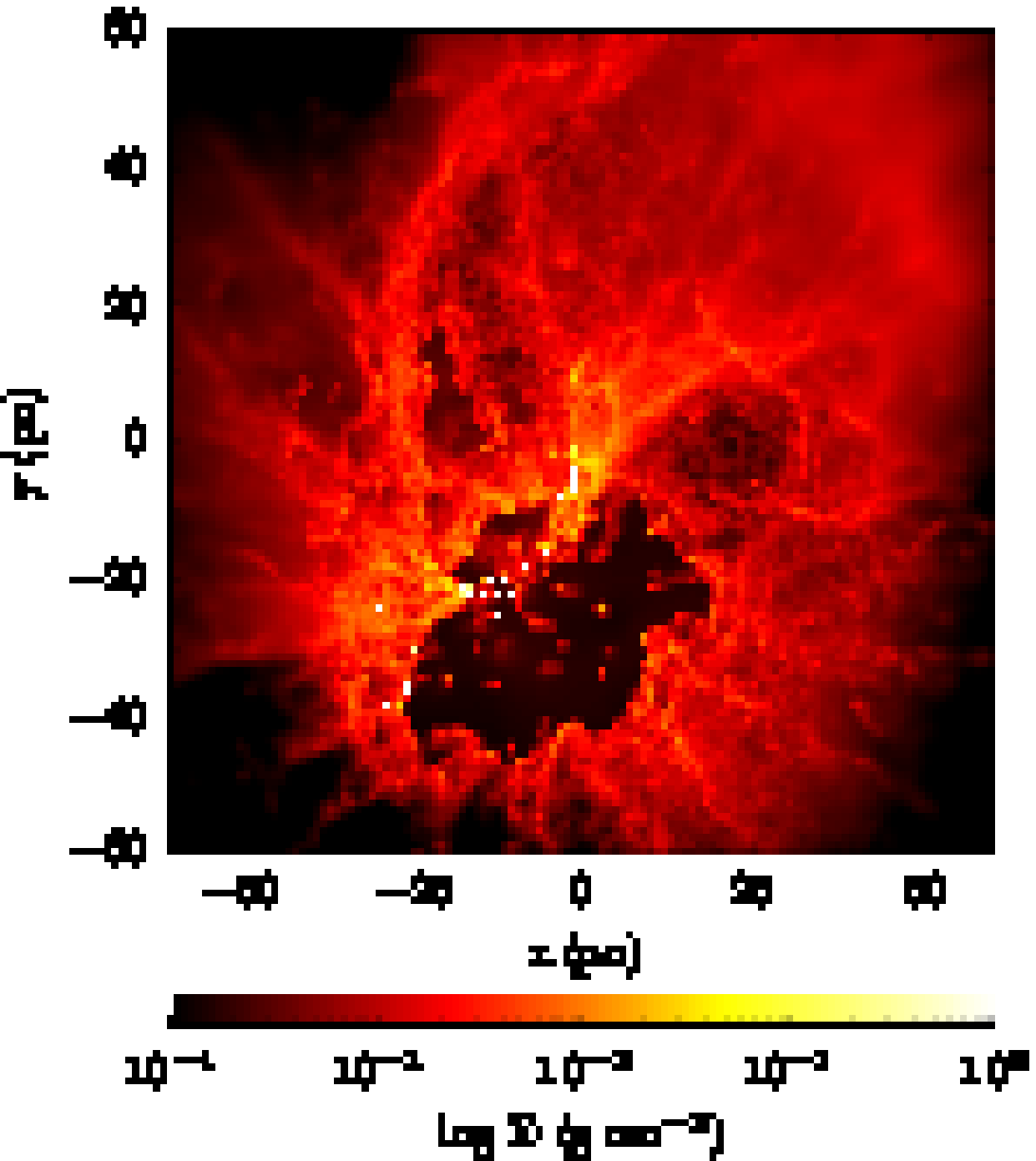}}
     \hspace{.1in}
     \subfloat[Run E]{\includegraphics[width=0.30\textwidth]{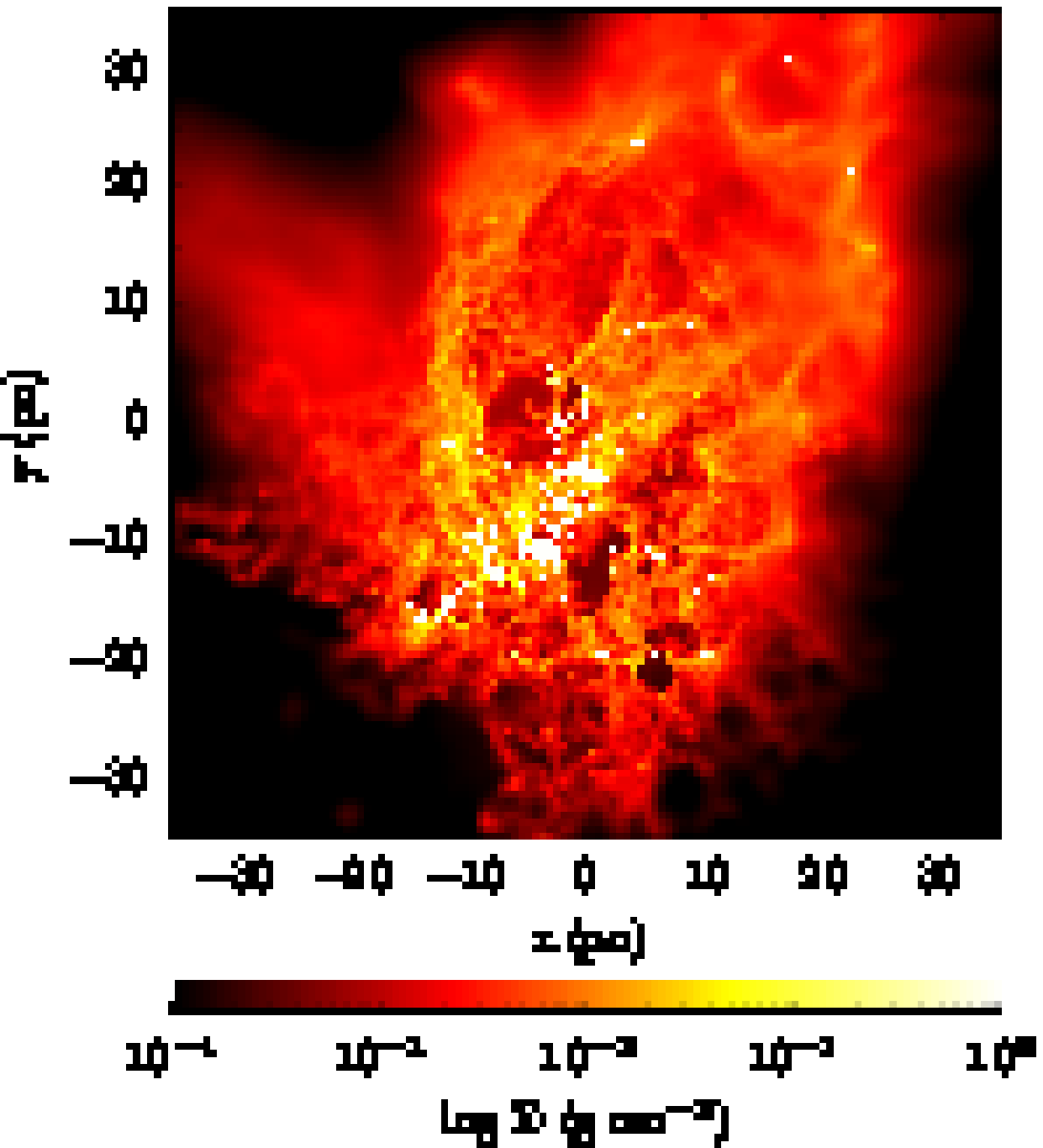}}
          \hspace{.1in}
     \subfloat[Run F]{\includegraphics[width=0.30\textwidth]{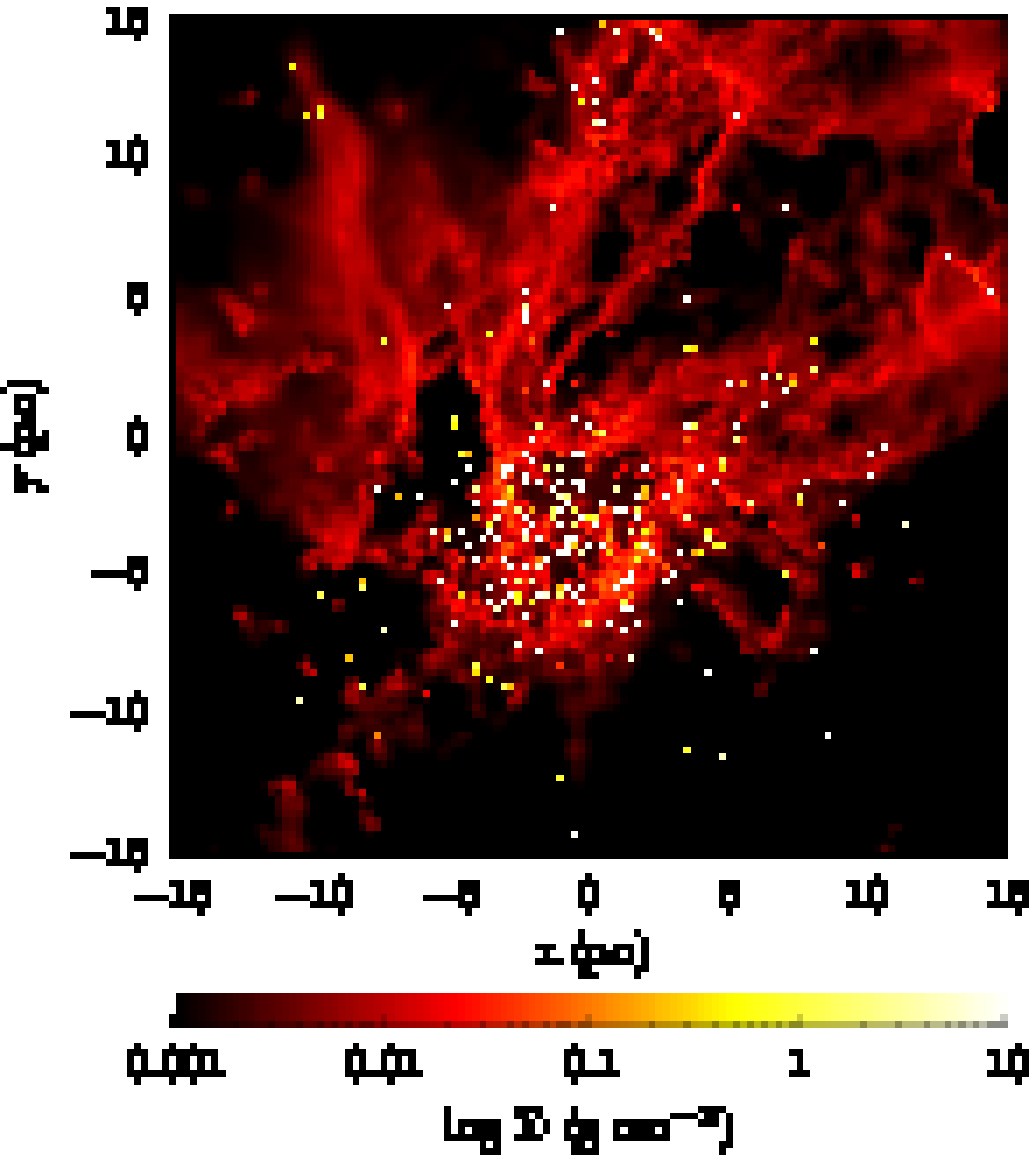}}
          \vspace{.1in}
     \subfloat[Run I]{\includegraphics[width=0.30\textwidth]{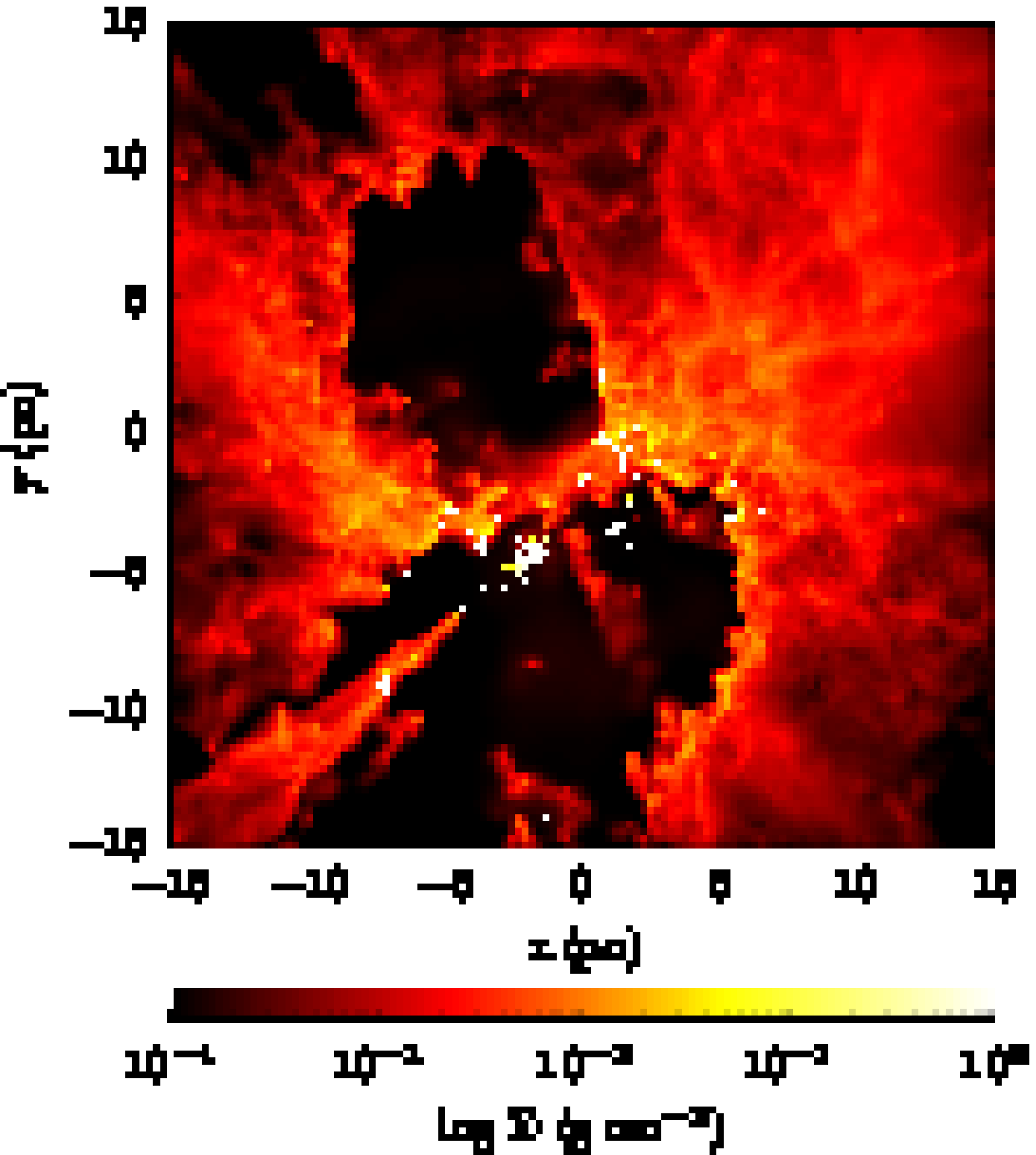}}
          \hspace{.1in}
     \subfloat[Run J]{\includegraphics[width=0.30\textwidth]{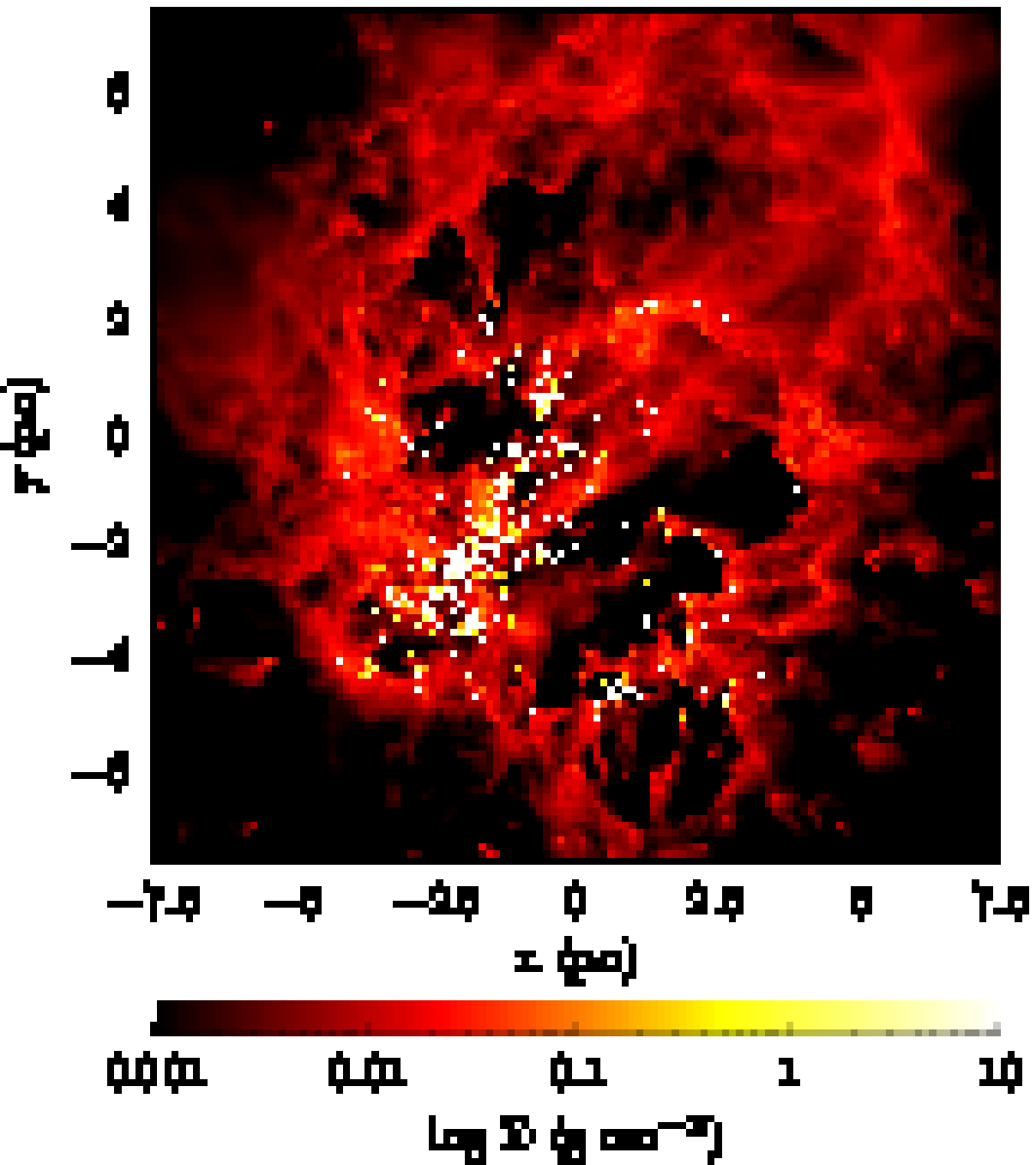}}
     \caption{Gallery of final states of clusters, as shown by column density maps observed down the $z$--axis. White dots represent sink particles (individual stars in Runs I and J, clusters otherwise) and are not to scale. Note the different physical sizes and the different column density scales.}
   \label{fig:gallery_final}
\end{figure*}   
\section{Effects of photoionization on model clusters}
The effects of photoionization in the $t_{\rm SN}$ time window before the first supernovae are expected to detonate vary strongly across our parameter space. There is a general gradient pointing from the denser massive, high--$v_{\rm RMS}$ clusters, to which it does little, towards the lower--density low--mass, low--$v_{\rm RMS}$ clusters, to which it is extremely disruptive. In Figure \ref{fig:gallery}, we show a gallery of the state of the clusters at the time when ionization is enabled. In Runs C, G and H, the implicit heating of the piecewise equation of state and the low turbulent velocities combine to prevent star formation. The high mean gas temperatures raise the thermal energy of these clouds so that clouds G and H are unbound and cloud C is nearly so. Furthermore, the low turbulent velocity dispersions and high temperatures result in mean Mach numbers so low that the turbulence dissipates without producing any very strong enhancements in the clouds' density fields. These clouds warm up and expand without forming any gravitationally--unstable structures.\\
\indent All the other clouds undergo star formation on timescales comparable to their freefall times. The impact of feedback varies strongly across the parameter space and we find that it is the lower--density star--forming clouds which are most strongly affected. Since these are also the warmer clouds, we repeated several simulations (Runs A, D, E and I) using an isothermal equation of state with a uniform temperature of 10K to evaluate the impact of the equation of state on our results. We find that this change makes little difference to the outcome of the simulations.\\
\indent All the systems have the same turbulent velocity field but different $v_{\rm RMS}$, so that many structures in the gas are visible in all simulations. All the systems exhibit a filamentary structure, with star formation being largely confined to the filaments and particularly to junctions connecting several filaments. This morphology is very similar to that observed increasingly frequently in cold dust emission by Herschel \citep[e.g.][]{2010A&A...518L.102A,2010A&A...518L.103M,2011A&A...533A..94H}. However, it is clear that the filamentary features in the gas are more sharply defined and complex in the systems with higher $v_{\rm RMS}$, higher mean densities and lower average temperatures (e.g. Runs B, X and F) than in the systems with lower Mach numbers and densities and warmer temperatures (e.g. Runs D, I). It is also clear that star formation in the low--density, low--Mach number systems is sparse, whereas in the denser and more strongly turbulent clusters such as Runs B, X and F, star formation has been more vigorous and evenly distributed. These are consequences of the higher initial average densities and the higher Mach numbers and stronger shocks in the latter systems.\\
\indent The subsequent behaviours and morphologies of the clouds are sufficiently diverse that we describe the reaction of each cluster in some detail below. The principal qualitative results may be gleaned from Figure \ref{fig:unbnd}, in which we examine the dynamical reaction of the clouds by plotting their star formation efficiencies, ionization fractions and unbound mass fractions (defined as the fraction of gas with positive total energy, including kinetic, thermal and gravitational components), comparing where appropriate with companion isothermal calculations (shown as dashed lines), and in Figure \ref{fig:gallery_final} where we show the final states of our calculations, in most cases after ionization has been acting for $\sim3$Myr.\\
\subsection{Run A (mass=$10^{6}$M$_{\odot}$, radius=180pc)}
This is the largest and most diffuse of our star--forming calculations (since neither Run C nor Run G nor Run H form any stars). The system forms a few tens of widely--separated clusters connected by filaments of denser gas from which they continually accrete (Figure \ref{fig:gallery}). As shown in Figure \ref{fig:gallery_final}, photoionization is able to partially disrupt the accretion filaments and creates a network of bubbles, several of which expand outwards from the cloud, becoming champagne flows [an example is visible at about  (-20,-150)]. The HII regions rapidly join up with one another so that large volumes of ionized gas are being illuminated by several sources. However, much of the volume of the cloud remains untouched by HII gas because the clusters are few and the cloud is very large. Consequently, as we show in Figure \ref{fig:unbnd}, the ionized and unbound gas fractions grow slowly and the effect on star formation, apparent from the flattening of the stellar mass curve, is slight but negative. The evolution of the companion isothermal calculation is very similar. Feedback begins acting earlier owing to the earlier formation of massive clusters in the isothermal calculation, but the quantities of ionized and unbound material extant after 3Myr of photoionization are very similar, although slightly lower.
\subsection{Run B (mass=$10^{6}$M$_{\odot}$, radius=95pc)}
Star formation in Run B is more vigorous and somewhat less sparse than in Run A, with $\sim100$ clusters being formed by the epoch at which ionization was switched on, largely as a consequence of Run B being smaller and denser so that the freefall time is shorter, and the star formation rate per Myr is higher. Owing to the higher gas densities, most of the ionizing sources are swamped by accretion flows according to the criterion given in Equation \ref{eqn:resc} and the fraction of ionized gas grows more slowly than in Run A. In addition, the escape velocity of Run B is higher than in Run A. The fractions of ionized and unbound gas and are able to reach only a few percent at $t_{\rm SN}$ and the influence on star formation is negligible, as shown in Figure \ref{fig:unbnd}. The morphology of the gas and thus the appearance of the cluster are little affected by feedback, but some disruption of the dense filaments and clearing of material away from some of the clusters has occurred, visible in Figure \ref{fig:gallery_final}.
\subsection{Run X (mass=$10^{6}$M$_{\odot}$, radius=45pc)}
This simulation is equivalent to that detailed in \cite{2011MNRAS.414..321D}, the only difference being that we employ our new more accurate and better physically--motivated ionization code. The evolution of the system is very similar to that in the above paper. In Fig \ref{fig:unbnd} we compare the evolution of the ionized gas fractions, unbound mass fractions and stellar mass using the multiple--source ionization algorithm from Dale and Bonnell 2011 (dashed lines) with Run X (solid lines), in which we use the newer algorithm. The older algorithm produces a very slightly higher ionization fraction, but the difference between the runs is minimal, with the plots of stellar mass fraction being indistinguishable. We confirm that the rate of ionization is very slow due to the strong accretion flows interacting with all the ionizing clusters, with only $\sim1$ percent of the gas being ionized per Myr. The rate at which gas becomes unbound is even lower than in Run B because the escape velocity of the Run X cloud is higher and in fact exceeds the typical sound speed in the ionized gas (this is the only one of our simulations for which this is the case). The effect on the star formation efficiency as a function of time is very small and in Figure \ref{fig:gallery_final}, where only neutral gas is shown, there are no discernible morphological signposts of feedback at all.\\
\subsection{Run D (mass=$10^{5}$M$_{\odot}$, radius=45pc)}
Run D is a diffuse, low--$v_{\rm RMS}$ system similar to Run A but considerably smaller. It also forms a few tens of rather sparsely--distributed clusters connected by filamentary structures in the gas. Ionization has a much stronger effect on this system, owing to its smaller size and escape velocity, the clusters being distributed over a larger fraction of the system volume, and the weaker accretion flows depositing mass onto them. The bubbles excavated by feedback are of comparable size to those seen in Run A, but, as shown in Figure \ref{fig:gallery_final}, occupy a much larger fraction of the volume of the system. The growth of the ionization fraction and influence of feedback on the dynamics are consequently stronger. Figure \ref{fig:unbnd} shows that ionization is able to unbind more than 25 percent of the cluster's gas reserves within $t_{\rm SN}$. However, Figure \ref{fig:unbnd} also shows that the effect of feedback on the overall star formation rate is minimal. Star formation in the isothermal Run D calculation starts earlier and proceeds to higher efficiencies than in the calculation using the Larson equation of state. Feedback therefore also begins acting acting earlier in the cloud's evolution. However, the fractions of ionized and unbound material resulting from 3Myr of photoionization in the two calculations are comparable, although somewhat lower in the isothermal case.\\
\subsection{Run E (mass=$10^{5}$M$_{\odot}$, radius=21pc)}
Run E is similar in structure to Run B at the onset of ionization and, from a dynamical point of view, the evolution of the two systems is similar, although somewhat more material is ionized and unbound in Run E (Figure \ref{fig:unbnd}. In neither cluster does ionization have any noticeable effect on the star formation rate. However, Figure \ref{fig:gallery_final} shows that photoionization influences the morphology of Run E significantly and is beginning to clear the gas away from the star clusters when the simulation was terminated after 2.3Myr of feedback -- there is a very clearly--defined bipolar bubble visible at (7,-20). Run E bridges the gap between the systems towards the high--mass, high--density corner of our parameter space on which feedback has no discernible effect, and the low--mass, low-density corner in which feedback profoundly alters the appearance and dynamics of the clouds. The companion isothermal calculation to Run E is almost identical, with slightly higher star formation efficiency, and slightly lower fractions of ionized and unbound gas.\\
\subsection{Run F (mass=$10^{5}$M$_{\odot}$, radius=10pc)}
Run F is a small, dense and high--$v_{\rm RMS}$ system which, in common with Run J, is unusual in the parameter space in that its freefall time is considerably shorter, at $\sim0.8$Myr, than the critical 3 Myr time window, so that the gravity--driven evolution of Run F proceeds very fast in comparison to this timescale. The effect of feedback on this cloud is initially very similar to that of Run X in that ionization has minimal impact. The ionized and unbound mass fractions grow very slowly and star formation proceeds largely unrestrained. However, at a system age of $\sim 4$Myr ($\sim 5 $t$_{\rm ff}$), when the star formation efficiency reaches and exceeds 60 percent, star formation begins to tail off and the ionization fraction begins to grow faster (although the unbound mass fractions evolves largely as before). This change in behaviour is due to the system beginning to run out of gas, clearly illustrated in Figure \ref{fig:gallery_final}, but because of star formation and accretion, not because of feedback.\\
\subsection{Run I (mass=$10^{4}$M$_{\odot}$, radius=10pc)}
Run I has the lowest turbulent Mach number of any of the star--forming systems in this study and its initial filamentary structure is consequently the most poorly--defined (Figure \ref{fig:gallery}) and its density field the smoothest. It also has the lowest escape velocity. As shown in Figures \ref{fig:unbnd} and \ref{fig:gallery_final}, Run I is the system on which feedback has the most dramatic effect, both dynamically and morphologically. Ionization excavates a vast cavity which appears roughly bipolar in shape when viewed along the $z$--axis, leaving two separated central clusters entirely devoid of gas, and creating several pillars pointing towards the most massive of the two. After 2.2 Myr, $\sim 58$ percent of the gas/stars have been unbound, although the ionization fraction remains modest at $\sim 10$ percent. Extrapolating the evolution of the ionization fraction, we conclude that $\sim 65$ percent of the system will be unbound by photoionization before the detonation of the first supernova. The effect on the star formation rate is small, although the star formation rate in this system is in any case low. Once again, the isothermal companion run exhibits very similar behaviour. Star formation and feedback both start $\sim$1Myr earlier in the isothermal run, but the evolution of the ionized and unbound gas fractions mirrors that in the standard run, although note that the isothermal calculation was only continued for $\approx$1.9Myr. As in all the other isothermal simulations, the fractions of ionized and unbound gas are slightly lower than in the standard run.\\
\subsection{Run J (mass=$10^{4}$M$_{\odot}$, radius=5pc)}
Run J in Figure \ref{fig:gallery} appears similar to Run I but with more star formation, owing to its shorter freefall time, and more well--defined filamentary morphology, due to the higher Mach number of the turbulence. The evolution of Run J is also strongly influenced by feedback. Figure \ref{fig:gallery_final} shows (after only 1.3Myr) a complex morphology, featuring several poorly--defined bubbles and pillar--like structures. Champagne flows were also observed at earlier stages in the evolution. Although we were only able able to evolve the system for $\sim1.3$Myr owing to the large number of stars formed ($\sim$600) Figure \ref{fig:unbnd} reveals that ionization has already unbound over 18 percent of the gas by this epoch.\\
\section{Discussion}
\begin{figure}
\includegraphics[width=0.5\textwidth]{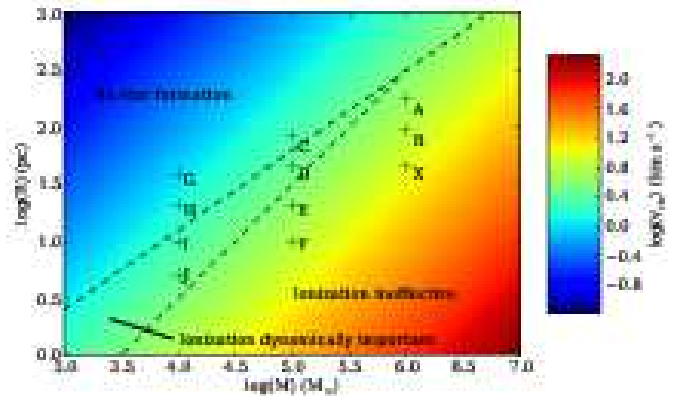}
\caption{The mass--radius parameter space studied in this work with colours representing cloud turbulent velocities and dashed lines partitioning the mass--radius plane according to whether star formation occurs and whether ionizing feedback is dynamically important.} 
\label{fig:result}
\end{figure}

Our principal goal in this paper was to determine whether photoionization from populations of O--stars or O--star--hosting clusters self--consistently formed within turbulent clouds covering a realistic parameter space could disrupt, or at least strongly dynamically influence the evolution of the clouds before the explosion of the first supernovae. We find that the answer to this question is yes, and our results are summarised in Figure \ref{fig:result} where we partition our parameter space according to which of our model clouds formed stars and of those, which are dynamically influenced by feedback. In the upper region of the plot, populated by Runs C, G and H, our chosen equation of state and assumptions about the turbulent velocity dispersion produce clouds whose velocity fields dissipate, and the clouds expand and evaporate, without forming any gravitationally--unstable structures. This result is in agreement with the conclusions of \cite{2011ApJ...731...25K} who used coupled thermal and chemical models of the ISM to determine the expected correlation between the masses of different ISM phases in a given galaxy, and the total star formation rate. They found that the star formation rate correlates most strongly with the mass of molecular gas, i.e. that most star formation occurs in that phase, and that atomic gas rarely forms stars. This result is also in line with the conclusions of \cite{2012MNRAS.tmp.2154G} who performed detailed thermal and chemical modelling of the ISM to determine whether molecular gas is a prerequisite for star formation to occur. They find that it is not -- it is the ability of clouds to self--shield against the external ISM radiation field that determines whether or not they will form stars. Clouds with low surface densities become too warm to form either molecules or stars. This agreement is perhaps not surprising, since \cite{2012MNRAS.tmp.2154G} recovered a temperature--density relation very similar to our assumed equation of state.\\
\indent The lower portion of Figure \ref{fig:result} containing all our other runs, contains clouds that form stars under the assumptions stated above. If we define the unbinding of $>$10 percent of the system's mass as constituting a strong dynamical effect, we may divide the lower region in two. The lowest part of the diagram, comprising Runs A, B, X, E and F are clusters on which ionizing feedback has little influence within $t_{\rm SN}$. In the triangular region in between, represented by Runs D, I and J, ionization is able to expel several tens of percent of the clouds' gas reserves before any supernovae detonate.\\
\subsection{The dynamical impact of ionization}
\indent The different reaction of the clusters to ionization is a consequence of several factors, as detailed in Section 4. In our simulations, the number and luminosity of the ionizing sources plays a minor role inasmuch as it is overwhelmed by other factors. The total ionizing luminosity at the end of Run X is a factor of approximately ten times higher than at the corresponding time in Run A but ionization has a much stronger effect on Run A. We also show explicitly in the Appendix that increasing all ionizing luminosities by a factor of two, or allowing the action of ionization to being earlier when there are fewer sources present have only minor impact on the outcome of a given simulation. The most important factors controlling the evolution of the clouds are their density fields, resulting in turn from their mean densities and turbulent velocity fields, and their escape velocities. The density fields influence photoionization via the accretion rates of the ionizing sources and the recombination rates in the surrounding gas. Using either or of the criteria expressed in Equations \ref{eqn:mdotcrit} or \ref{eqn:resc}, the growth of most of the HII regions in Runs X and F, for example, are swamped by accretion -- the ionizing fluxes of the sources are unable to keep pace with the inflows of neutral gas delivered by accretion flows -- whereas this is not true for any ionizing sources in Runs A, D, I or J. However, Figure \ref{fig:unbnd} indicates that this is not the most important factor controlling the dynamical influence of ionization. In most of the calculations, the total ionized fraction of the cloud after 3Myr is in the range 3-10$\%$, yet the unbound mass fractions range from $\sim1\%$-$\sim60\%$. Runs E and J, for example, have very similar ionization fractions in their final states of $\sim5\%$ but three times more mass has been unbound in Run J. It is clear that the ability of ionized gas to entrain and expel neutral gas varies greatly across the parameter space, and this is governed by the clouds' escape velocities. Since the escape velocity depends on the ratio $M/R$ and the clouds in this study and the \cite{2009ApJ...699.1092H} work follow approximately $M\propto R^{2}$, it follows that larger clouds (in either sense) have higher escape velocities and are thus intrinsically more difficult to unbind with photoionization. It is this factor which is most important in determining how much damage ionization can do to a given cloud.\\
\indent We note that, in \textit{none} of our model clusters, even those in which large fractions of the neutral gas were swept up and expelled by ionization, is the star formation rate significantly altered in either a positive or negative sense. In the denser higher--mass systems such as Runs B, X and F, this is largely due to the fact that ionization has very little effect on the cloud dynamics, so that the star formation process feels no effect. In contrast, in the lower density, lower mass  clusters which are strongly influenced by feedback, it appears that the negative impact of losing large fractions of the clouds' neutral gas is roughly compensated for on the timescales of these simulations, by positive feedback in the form of triggered star formation. Runs D and I in particular exhibit several features generally associated with triggering such as pillars with stars at their tips and stars embedded in ridges of dense gas around the edges of feedback--driven bubbles. We defer a detailed study of triggering in these simulations to a later paper.\\
\subsection{Influence of the equation of state}
\begin{figure}
\includegraphics[width=0.5\textwidth]{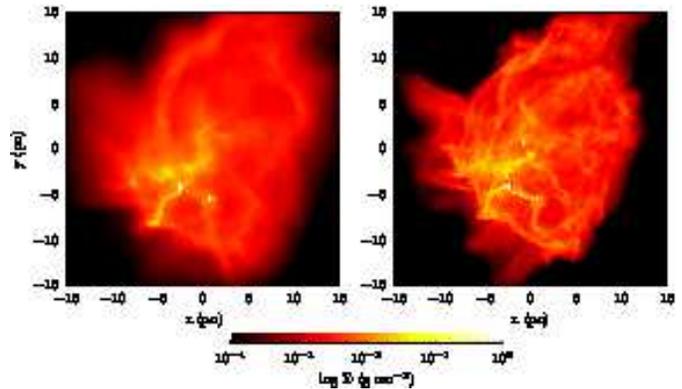}
\caption{Comparison of the states of the standard (left panel) and isothermal (right panel) Run I calculations at the point where feedback was enabled.} 
\label{fig:runi_iso_init}
\end{figure}

\begin{figure}
\includegraphics[width=0.5\textwidth]{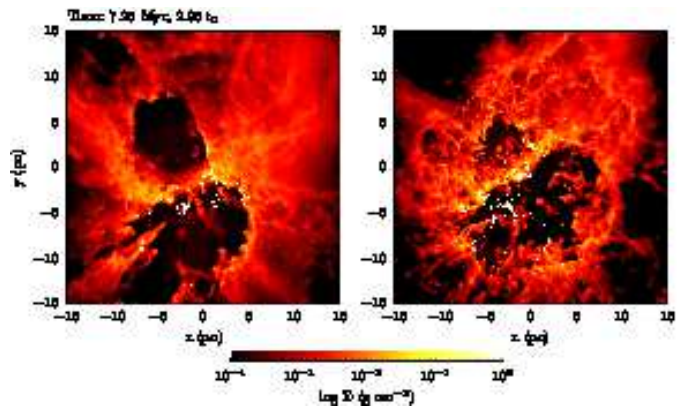}
\caption{Comparison of the states of the standard (left panel) and isothermal (right panel) Run I calculations after 1.9Myr of feedback.} 
\label{fig:runi_iso_final}
\end{figure}

In most of our calculations, we made use of the Larson equation of state, as detailed in Section 2. Owing to the large range in densities found in our parameter space, the use of this EOS resulted in some of our clouds being warm enough that they should really be regarded as at least partially atomic, rather than molecular, and we found that these clouds failed to form stars under the influence or our chosen turbulent velocity fields. However, several clouds, e.g. Runs A, D and I were still able to form stars despite their warm average temperatures. To quantify the impact of our assumed EOS, we repeated these simulations (as well as Run E) with an even simpler EOS, namely one with a fixed isothermal temperature of 10K. In Figures \ref{fig:runi_iso_init} and \ref{fig:runi_iso_final} we show column--density maps of the states of the warm (left panels) and isothermal (right panels) Run I clouds at the time feedback was enabled and after $\sim1.9$Myr of feedback respectively. The influence of the equation of state on the \emph{morphological} evolution of the clouds without feedback is very clear from Figure \ref{fig:runi_iso_init}. There is more fine--scale structure in the isothermal cloud, as would be expected if the average gas temperature is cooler. The morphology of the star formation is rather similar, however, with the main concentration of stars being located at a junction of several filaments of dense gas at approximately (-3, -4)pc in both clouds. This is largely because the star formation forms in the densest coolest gas, which is treated isothermally in all simulations, and these regions are created by converging turbulent flows, which are also the same in all simulations. Turning to Figure \ref{fig:runi_iso_final}, the effects of feedback on the clouds are also similar. In both cases, the largest stellar concentration located at the filament junction is responsible for most of the ionizing luminosity and generates a roughly bipolar bubble morphology bisected by a ridge of dense material extending roughly diagonally from the lower left to the upper right (at least when viewed in this projection). The bubbles are less symmetrical in the isothermal case, with the upper bubble being much smaller than the lower. In addition, the larger bubble in the isothermal calculation appears to be more spherical in shape and less well cleared out, although this is partly a projection effect. The initially smoother gas in the warm cloud has therefore led to a somewhat simpler bubble morphology. The distribution of stars in the two calculations is also very similar, with both possessing a few well--defined clusters and more distributed star formation in the ridges of dense material.\\
\indent However, as shown in Section 5 above, the gross dynamical behaviour of these two simulations is very similar. The temperature of the gas has very little impact on the influence of ionizing feedback. The higher average gas densities in the isothermal calculation result in star formation and feedback beginning earlier but does not much affect the star formation rate or efficiency. This is probably because the isothermal and Larson versions of each simulation have identical turbulent velocity fields and gravitational wells, which between them are what generate the cold gas from which the stars form. The earlier star formation in the isothermal calculations is a result of all the gas being already cool and the shocks being consequently somewhat stronger, but differences in the mean initial gas temperature, unless large enough to make the clouds nearly or actually unbound, do not strongly affect the rate at which cold, dense gas is generated. The mean gas densities in the isothermal calculations are slightly higher than in the standard runs, resulting in a somewhat lower fraction of gas being ionized or unbound at a given time but, as we have seen in all calculations, the depth of the cloud potential well is a much stronger determinant of the impact of feedback.
\subsection{Escape of ionizing photons}
\indent The source of photons required to maintain the diffuse ionized gas (DIG) layers observed above and below galactic disks is still under discussion. On energetic grounds \citep[e.g.][]{1984ApJ...282..191R}, OB stars are strong candidates, but the thickness of such ionized layers is often much greater than that of the thin galactic disks where O--stars are likely to be found. In order for O--stars to accomplish this feat, significant fractions of their photon fluxes must be able to escape their natal molecular clouds, or the O--stars themselves must escape the clouds. \cite{2000ApJ...541..597H} attempted to reproduce the DIG characteristics of M33 and concluded that field O--stars could not provide sufficient photons. Both they and \cite{2006ApJ...644L..29V} calculated that the DIG in M33 could be explained by a combination of field O--stars and HII regions leaking $\sim$30$\%$ of their photons into the ISM. There are additional problems with the O--star photoionisation model, principally the degree of radiation hardening required to reproduce the observed line ratios \citep[e.g.][]{2004MNRAS.353.1126W}, which are beyond the scope of this paper. However, if OB stars are to be responsible for ionizing the DIG and field O stars cannot do the job alone, GMCs and the HII regions they host must have a high degree of porosity. Figure \ref{fig:gallery_final} suggests that several of our model clouds, particularly those on which feedback has had a strong influence, may indeed allow large fractions of their ionizing luminosities to escape.\\
\indent We estimate the escape fractions of ionizing photons at the ends of our simulations as follows: We first allow the multiple--source ionization code to iterate to a solution for the number of ionized particles and the photon fluxes reaching all particles.  We then loop through all the ionizing sources, construct a Hammer spherical grid centred on each one, locate the most distant SPH particle in each angular grid cell and compute the flux emerging radially from that particle (if any). We then sum all the emergent fluxes for all sources to compute the total emergent photon flux. In Table \ref{tab:photons}, we tabulate the total flux of all sources in each simulation, the fraction of these ionizing photons that are escaping f$_{\rm phot}$ and the total flux of photons escaping each cloud, $Q_{\rm H}$'.\\
\begin{table*}
\begin{tabular}{|l|l|l|l|}
Run & Total $Q_{\rm H}$& f$_{\rm phot}$ & $Q_{\rm H}$' (10$^{49}$ s$^{-1}$)\\
\hline
A&6.2&0.21&1.3\\
B&11.2&0.17&1.9\\
X&38.6&0.07&2.7\\
D&2.3&0.72&1.7\\
E&4.1&0.61&2.5\\
F&18.0&0.88&15.8\\
I&2.6&0.90&2.3\\
J&2.2&0.80&1.8\\
\end{tabular}
\caption{Total ionizing photon fluxes, estimated fraction of ionizing photons leaking from clouds and ionizing photon luminosities leaking from each cloud.}
\label{tab:photons}
\end{table*}
\indent The escape fraction varies widely and broadly in line with how much damage ionization has done to each cloud. With the exception of Run F, all the clouds' apparent ionizing luminosities lie in a very small range, because clouds with the highest actual luminosities have the lowest escape fractions and vice versa. Run F does not follow this trend because alone among the simulated clouds, star formation in Run F has almost gone to completion, so the cloud has both a large number of stars and little remaining gas to absorb their photons. We make no attempt to address the problem of explaining the large observed line ratios in the DIG, but our results do suggest that leakage from HII regions can at least provide sufficient photons.\\
\subsection{Paving the way for supernovae}
\indent The simulations presented here have been evolved as close as possible to 3Myr after the onset of photoionization on the grounds that this is the approximate main sequence lifetime of the most massive stars, and so that the isolated effect of photoionization may be observed. It is clear from Figure \ref{fig:gallery_final} that the morphology of many of the simulated clouds has been profoundly altered by the action of radiative feedback on this timescale, so that the environment in which the first supernovae in each cloud detonates has also been affected. A proper understanding of the consequences of this fact requires the simulation of the supernovae themselves, which we defer to later work. The problem in reality is also complicated by the action of winds, which we have neglected here. However, we may gain a crude idea of the impact of subsequent supernovae on our clouds by considering the fate of the momentum (which, unlike energy, cannot leak or be radiated away) injected by each explosion. We take each O--star or O--star hosting cluster in our simulations to explode spherically--symmetrically, ejecting a mass $M_{\rm EJECTA}$ with a total initial kinetic energy $E_{0}$, and construct a spherical grid around the explosion site. Each radial bin subtends a solid angle $d\Omega$, contains a mass of cloud materail $M(\theta,\phi)$ and absorbs a fraction $d\Omega/4\pi$ of the emitted momentum. We then compute by conserving momentum whether, in each radial bin, the final velocity of the combined ejecta and swept--up mass exceeds the cluster escape velocity. This then gives crude estimates of (i) what fraction f$_{\rm esc}$ of the ejecta escapes the cluster (ii) what fraction f$_{\rm stop}$ of the ejecta is slowed down sufficiently to be involved in further star formation (iii) how much mass M$_{\rm unbnd}$ is likely to be unbound by the explosion (treating each supernova as a separate event and neglecting cumulative effects). If we take $M_{\rm EJECTA}$=10M$_{\odot}$ and $E_{0}$=10$^{51}$ erg, we obtain the values given in Table \ref{tab:sn}.\\
\indent The interpretation of this crude model should not be taken too far, but Table \ref{tab:sn} implies that the porosity of the clouds to supernova ejecta is little affected by ionizing feedback and is instead simply related to the clouds' masses and boundedness. This impression is reinforced by repeating the analysis on the clouds \textit{before} photoionization is switched on, which produces escape fractions only $\sim5\%$ smaller than those given above. The results also suggest that most supernova ejecta in the 10$^{5}$ and 10$^{6}$ M$_{\odot}$ clouds will be retained by the clouds and involved in further star formation, and that the clouds will survive several supernovae. The lower mass clouds, in contrast, will lose most of their supernova ejecta but a small fraction of the original cloud mass $\sim20\%$ is likely to survive both ionization feedback and the first supernova and may host star formation involving chemically--enriched material.\\
\subsection{Other forms of feedback}
\indent We have considered only the effects of photoionization in this work but in reality, other forms of feedback will act contemporaneously in star--forming regions, even before the detonation of the first supernovae, and these other mechanisms may help or hinder each other in unbinding clouds and expelling material. Winds (spherical outflows) and jets (collimated outflows) also inject energy and momentum into the circumstellar gas. Although jets have lower velocities, they have higher mass--loss rates per star than main--sequence winds and their contribution to clouds' momentum budgets is comparable to or greater than that of winds, but much less than that of HII regions, as shown by \citep{2002ApJ...566..302M}. HII regions are likely to be even more dominant in low--metallicity systems where stellar winds (but not jets) are weak. Numerical simulations by \cite{2006ApJ...640L.187L} and analytical work by \cite{2007ApJ...659.1394M} found that multiple protostellar jets were able to maintain the supersonic turbulent velocity fields in clouds, although \cite{2007ApJ...668.1028B} arrived at the opposite conclusion from their numerical study. It is not clear, however, whether winds and jets acting together with photoionization will have a greater or lesser effect than any of these mechanisms acting alone. Winds or jets may destroy dense circumstellar material close to ionizing sources, allowing photons to penetrate further into a cloud, but winds may also sweep up low--density material into dense shells which confine HII regions. Conversely, jets are likely to punch holes in any shell--like structures, allowing ionizing photons and hot gas to escape. In their radiation--hydrodynamics simulations of protostellar disks with jets, \cite{2011ApJ...740..107C} observed just this phenomenon. Although they themselves make minor contributions to the momentum budget, winds and outflows may potentially alter the effectiveness of the the major contributor -- the HII regions -- but it is not easy to say by how much, or even in which direction.\\
\indent This question is further complicated by the action of magnetic fields, which we also neglect here. \cite{2007ApJ...671..518K} and \cite{2012ApJ...745..158G} simulated the evolution of HII regions in magnetized clouds and found that the energy uptake by the gas was much more efficient than in the unmagnetized case. \cite{2010ApJ...709...27W} found a similar result when comparing simulations including protostellar outflows in the presence and absence of magnetic fields. The magnetic field increases the coherence of outflows and enhances their ability to sweep up material, thereby making them more efficient at expelling circumstellar gas. In the case of HII regions, the efficiency of cloud destruction is still potentially limited by the sound speed in the HII region in the case that the escape velocity of the cloud is comparable or higher. The additional magnetic pressure may lower the effective escape velocity, but the field needs to be rather strong for this to be a large effect.\\
\begin{table*}
\begin{tabular}{|l|l|l|l|}
Run & f$_{\rm esc}$ & f$_{\rm stop}$ & M$_{\rm unbnd}$ (M$_{\odot}$)\\
\hline
A&$<$0.01&$>$0.99 & 190\\
B&$<$0.01&$>$0.99 & 180\\
X&$<$0.01&$>$0.99 & 150\\
D&0.10& 0.90 & 320\\
E&0.20& 0.80 & 320\\
F&0.10& 0.90 & 60\\
I&0.70& 0.30 &1990\\
J&0.65& 0.35 &1380\\
\end{tabular}
\caption{Estimated effects of single supernovae on clouds after the action of photoionization.}
\label{tab:sn}
\end{table*}
\section{Conclusions}
Ionizing feedback from O--type stars can have a strong dynamical effect on embedded clusters on the 3Myr timescale before the first supernovae detonate. The influence of feedback is stifled in clouds with higher densities and more sharply--defined structure by the swamping of HII regions by accretion flows. More importantly, clouds with higher escape velocities are more resistant to feedback, since the maximum expansion speed of an HII region is $\sim10$ km s$^{-1}$. Since observed clouds (and those modelled here) follow approximately the relation $M\propto R^{2}$, more massive and larger clouds are intrinsically more difficult to unbind. These two factors contribute to making the effects of ionization feedback within the 3Myr window much stronger on the lower--density low mass (10$^{4}$M$_{\odot}$) clouds than on the densest high--mass 10$^{6}$M$_{\odot}$ clouds. That massive clouds with larger escape velocities are difficult or impossible to disrupt by thermal pressure from photoionized gas has been suggested before by \cite{2009ApJ...703.1352K}, \cite{2010ApJ...709..191M} and \cite{2010ApJ...710L.142F}, all of whom posit that in fact radiation pressure is a more important feedback mechanism in high--mass clouds. Our detailed hydrodynamic calculations demonstrate that this is indeed the case.\\
\indent The influence of ionizing feedback on the rate and efficiency of star formation was small in all our model clusters, either through the general inability of feedback to perturb the more massive and denser clouds, or through an approximate cancelling of disruption and triggering in the low--mass more diffuse systems, which we will explore in detail in a subsequent paper. Overall, the fact that stars form in the densest gas, often pre--existing thanks to the turbulence with which the clouds were seeded, limits the effect \\
\indent The lower--mass and lower--density clouds become highly porous to ionizing photons due to the influence of feedback, but overall (with the exception of the gas--deprived Run F), the effective ionizing luminosities of all the clouds are the same to within a factor of $\sim2$ because the more porous clouds are also those with the smaller numbers of O--stars. Given that low mass clouds are more common than high--mass ones, this results implies that most photons in the ISM which were produced by O--stars come from low--mass clouds.\\
\indent The low--mass clouds are also porous to supernova ejecta, but this is not much aided by the action of photoionization. 60--70$\%$ of the ejecta from the 10$^{4}$M$_{\odot}$ clouds should be returned directly to the ISM, while the rest is likely to become involved in a second round of star formation in the remains of the clouds. In the higher mass clouds, most of the supernova ejecta should become involved in further star formation.\\
\indent Morphologically, our simulations exhibit virtually every feature commonly associated with feedback. We see evacuated cavities with pillars containing young stellar objects pointing towards massive stars similar to those seen in the Eagle Nebula \citep[e.g.][]{2002ApJ...565L..25S}, bubbles reminiscent of those observed by \cite{2006ApJ...649..759C} and champagne flows like that detailed by \cite{2007MNRAS.379.1237M}. However, we note that these are qualitative comparisons based on our simplistic column--density plots. In a later work, we will perform artificial observations of our results in commonly--used emission lines such as O[III] and N[II] that may be more quantitatively compared with observations.\\
\indent The next steps in this work are to examine in detail the triggering of star formation within the clouds, to extend the parameter space to initially--unbound clouds, to include the effects of stellar winds and to simulate the effects of supernovae.\\
\section{Acknowledgements}
The authors thank the anonymous referee for careful and constructive criticism which has significantly improved the paper.\\

\bibliography{myrefs}

\section{Appendix}
In this section, we detail some additional simulations performed to test the influence of some of the assumptions made in our models. Most of these tests were conducted on Run E, since it lies in the middle of the parameter space, near the approximate demarcation line between systems that are and are not strongly affected by photoionization.
\subsection{Effect of increasing the ionizing luminosity}
\begin{figure}
\includegraphics[width=0.5\textwidth]{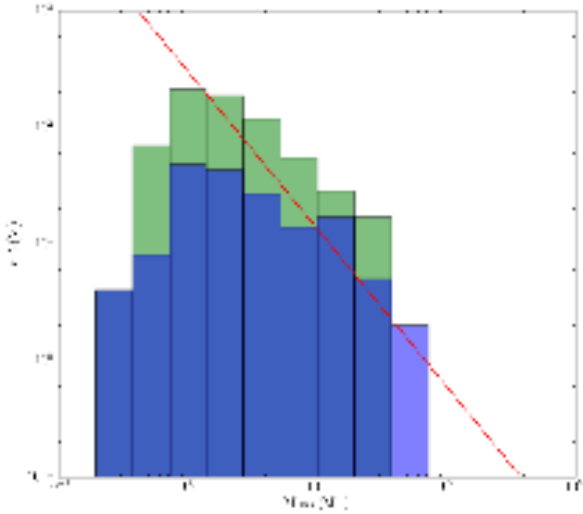}
\caption{Comparison of the stellar mass functions in Run I (blue) and Run J (green), with the Salpeter mass function (red line).} 
\label{fig:mfij}
\end{figure}

\begin{figure}
\includegraphics[width=0.5\textwidth]{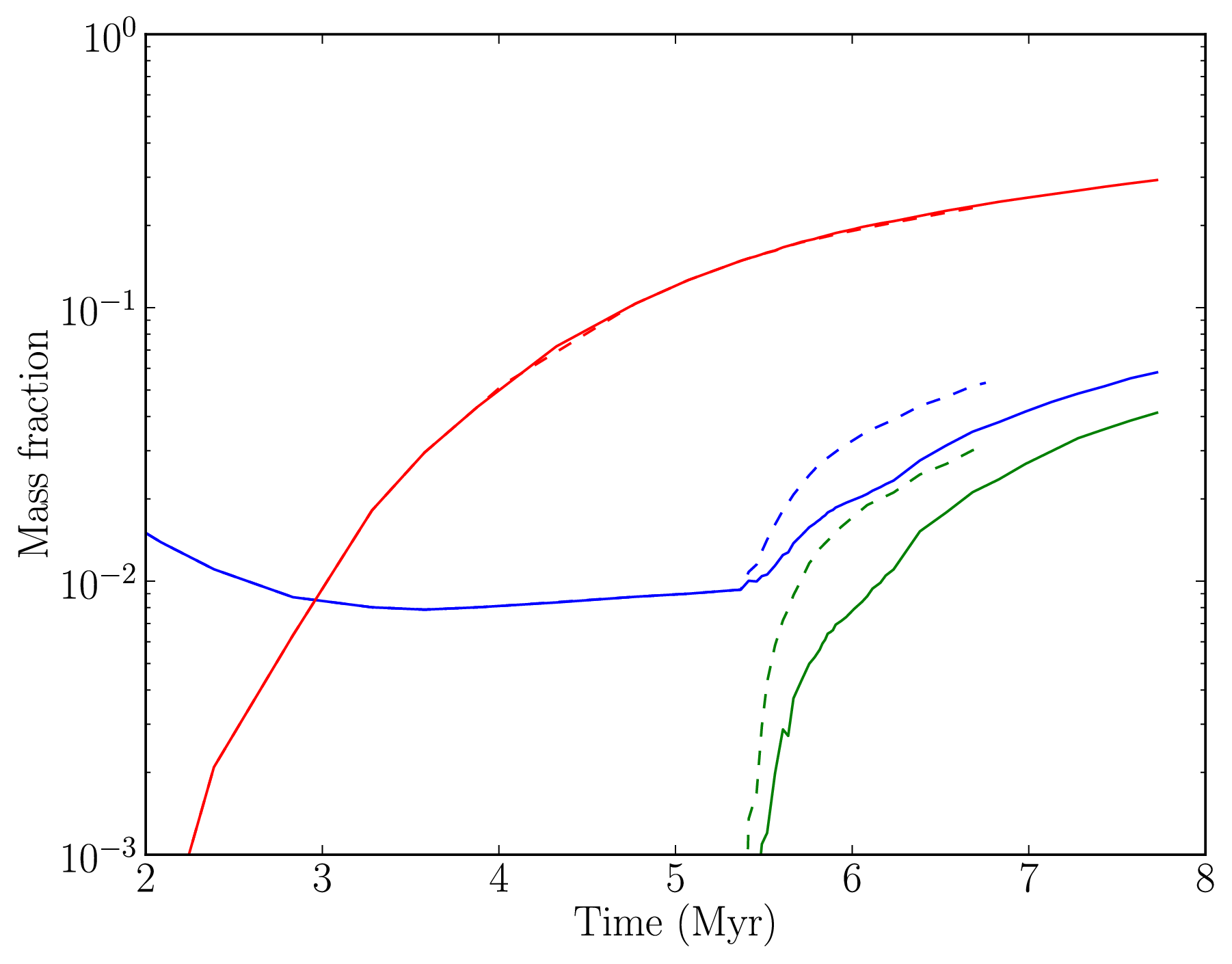}
\caption{Comparison of the evolution of the standard (solid lines) Run E with a duplicate run in which all ionizing photon fluxes are twice as big (dashed lines).} 
\label{fig:rune_2Q}
\end{figure}

In most of our simulations, the sink particles represent small clusters rather than individual stars and the ionizing luminosity must therefore be estimated from each cluster's mass using an IMF. This involves making assumptions about the IMF slope and lower--mass cutoff, which determines how much mass in a given cluster is locked up in OB--type stars, and computing from this mass an ionizing luminosity. Given that the most massive sink particles representing clusters, even in our $10^{6}$M$_{\odot}$ clouds, have masses of typically only a few thousand M$_{\odot}$, the upper ends of their IMFs will be poorly sampled and only a simple estimate of their ionizing luminosities is appropriate. However, in Runs I and J, we resolve individual stars and Run J has a reasonably well--sampled IMF containing $\sim 700$ objects by the end of that simulation. The total stellar mass at the end of Run J is $\sim2 400$M$_{\odot}$ and the total ionizing photon flux is $2.2\times10^{49}$s$^{-1}$. If the formula detailed in Section 2 is used to compute the photon flux of a 2400M$_{\odot}$ cluster, the result obtained is $0.9\times10^{49}$s$^{-1}$, or about half as luminous. To compute the luminosity of our unresolved clusters, we integrate over an assumed Salpeter IMF between mass limits of 0.1 and 100M$_{\odot}$ to compute the total mass in stars whose mass exceeds 30$M_{\odot}$, whereas the mass resolution in Run J is $\sim0.5-1$M$_{\odot}$. In Figure \ref{fig:mfij} we confirm that the mass functions in Runs I (blue) and J (green) are well fit by the Salpeter power law, depicted by the red line. If we were instead to use 0.5 or 1.0M$_{\odot}$ as the lower IMF mass limits, our unresolved clusters would be respectively 1.9--2.6 times more luminous. Taking a factor of two to be a reasonable uncertainty in our cluster photon fluxes, we therefore repeated Run E with the luminosities of all clusters increased by a factor of two. The result of the simulation is shown in Figure \ref{fig:rune_2Q}, compared to the standard computation with the original luminosities. We see that, initially, the quantity of ionized material in the calculation with enhanced luminosities is approximately a factor of two larger than in the standard run. This is to be expected, since the Str\"{o}mgren radius is proportional to $Q_{\rm H}^{1/3}$ and the volume of initially--ionized material is therefore proportional to $Q_{\rm H}$. However, the differences in both ionized and unbound mass between the two simulations do not grow but instead shrink with time and the overall evolution of the runs is very similar. Uncertainties of a factor of a few in the ionizing fluxes of sources are less important to the evolution of the clouds than the potential well in which the HII regions are trying to expand.
\subsection{Effect of switching ionization on earlier}
\begin{figure}
\includegraphics[width=0.5\textwidth]{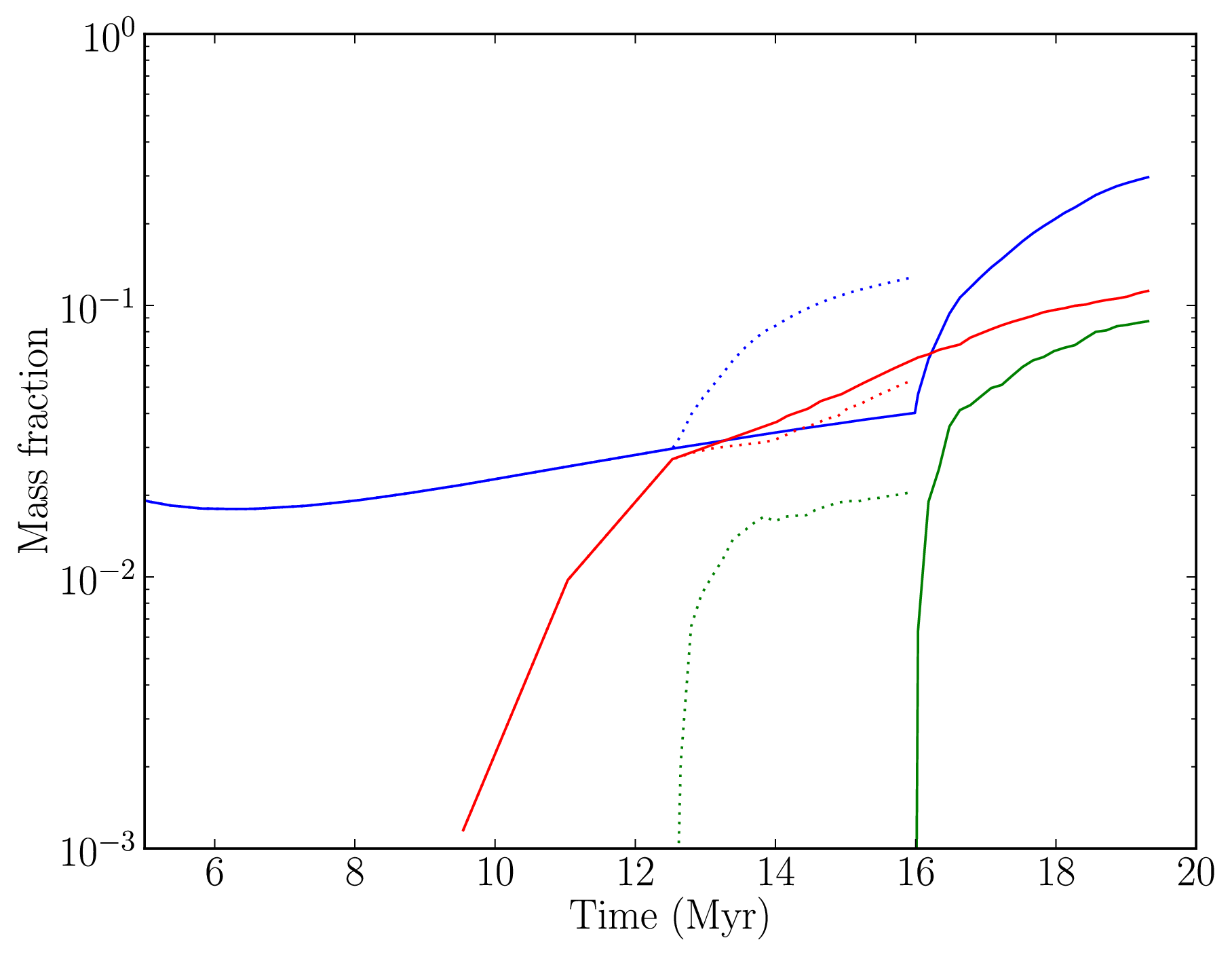}
\caption{Comparison of the evolution of the standard (solid lines) Run D with a duplicate in which ionization is enabled as soon as is possible (dashed lines).} 
\label{fig:rund_soon}
\end{figure}

\begin{figure}
\includegraphics[width=0.5\textwidth]{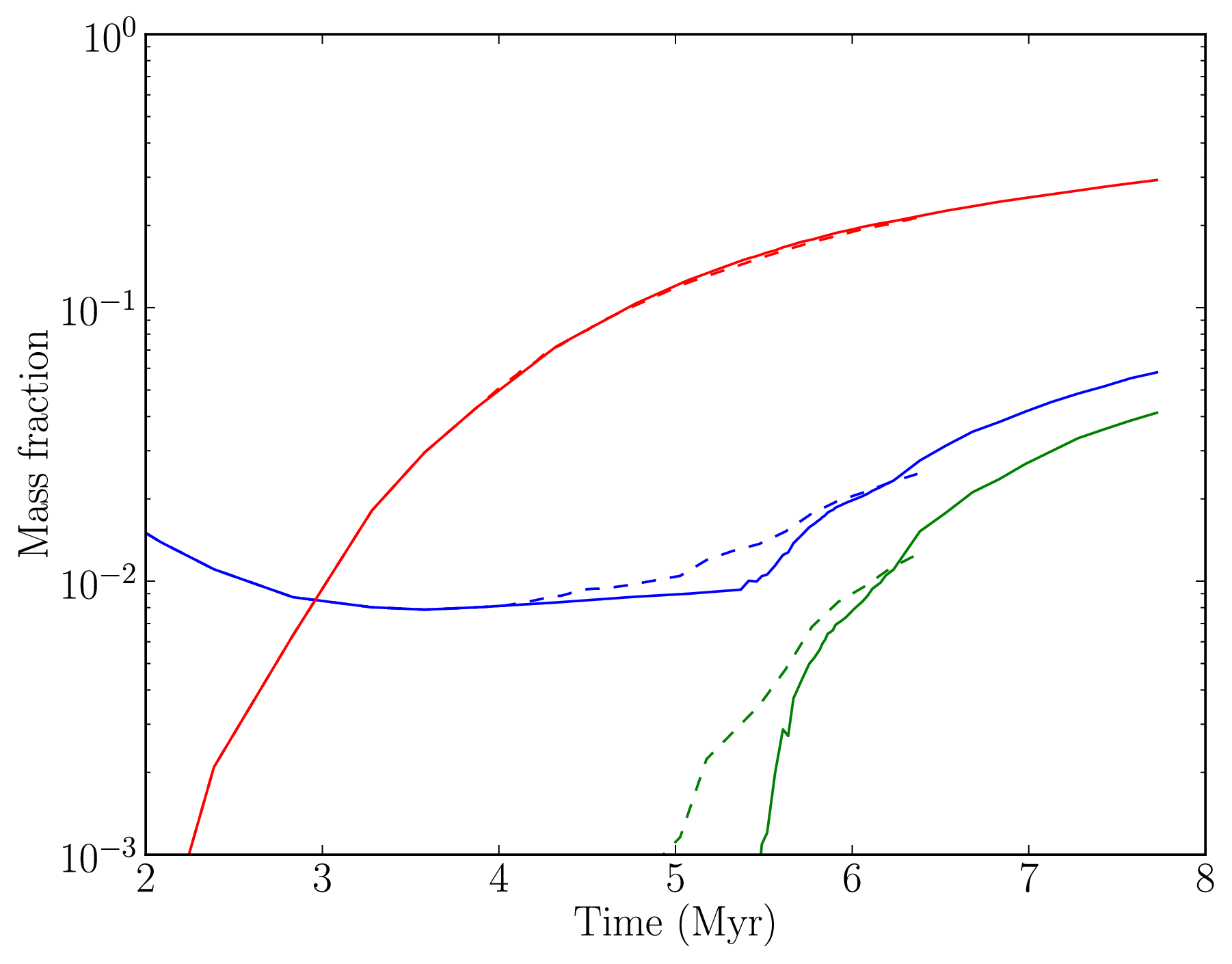}
\caption{Comparison of the evolution of the standard (solid lines) Run E with a duplicate in which ionization is enabled as soon as is possible (dashed lines).} 
\label{fig:rune_soon}
\end{figure}

This study aims at a better understanding of the large--scale impact of multiple ionizing sources on star--forming regions. The ionizing photon flux is dominated by the few most massive sources, and is therefore potentially subject to stochastic effects. To lessen these effects, we allowed our model clouds to develop a small number of sources massive enough to possess ionizing fluxes before enabling feedback. To quantify such effects, we repeated Runs D and E, enabling feedback as soon as the first object became sufficiently massive. The results are shown in Figures \ref{fig:rund_soon} and \ref{fig:rune_soon} compared with the standard runs in both cases. In the case of Run D, this results in feedback beginning $\sim$3Myr earlier in the system's evolution (equivalent to $\sim0.45$ freefall times). The result is a considerably smaller mass of material being ionized, reducing the global ionization fraction after 3Myr of feedback from $\sim9\%$ to $\sim2\%$. However, the difference in the amount of material unbound on the same timescale is only a factor of two -- $\sim15\%$ as opposed to $\sim30\%$. The impact on Run E is rather similar. Feedback is enabled $\sim1.5$Myr earlier ($\sim0.60$ freefall times), resulting in the ionization fraction after 2.3Myr of feedback being reduced by a factor of $\sim$four and the unbound gas fraction being reduced by a factor of $\sim$two. Exactly when feedback begins acting on a cloud clearly has some influence on the evolution but varying this time by substantial fractions of the system freefall time produces only modest differences in the quantities of material unbound.
\subsection{Effect of numerical resolution}
\begin{figure}
\includegraphics[width=0.5\textwidth]{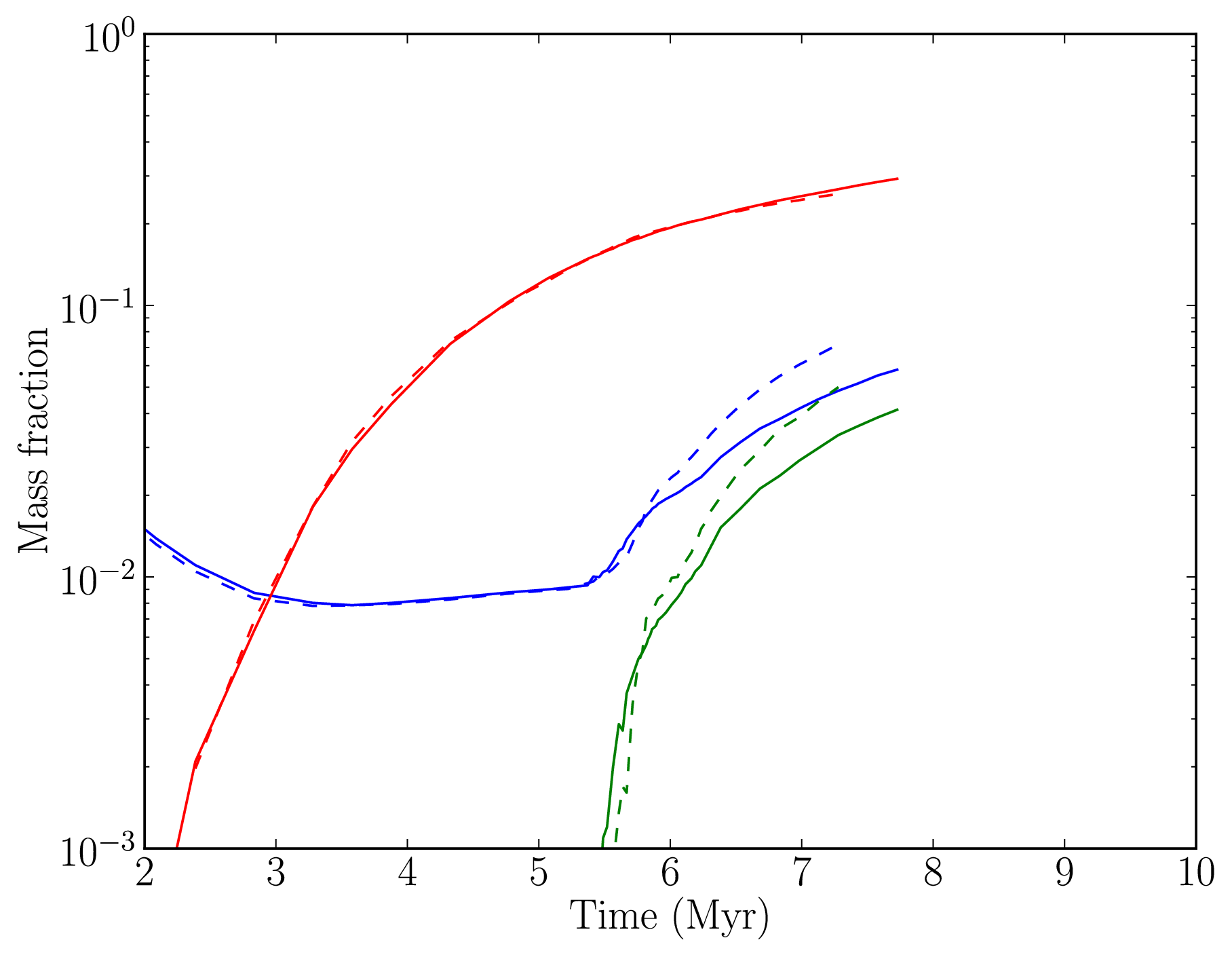}
\caption{Comparison of the evolution of the standard (solid lines) $10^{6}$--particle Run E with a duplicate low--resolution $5\times10^{5}$--particle run (dashed lines).} 
\label{fig:rune_res}
\end{figure}

\begin{figure}
\includegraphics[width=0.5\textwidth]{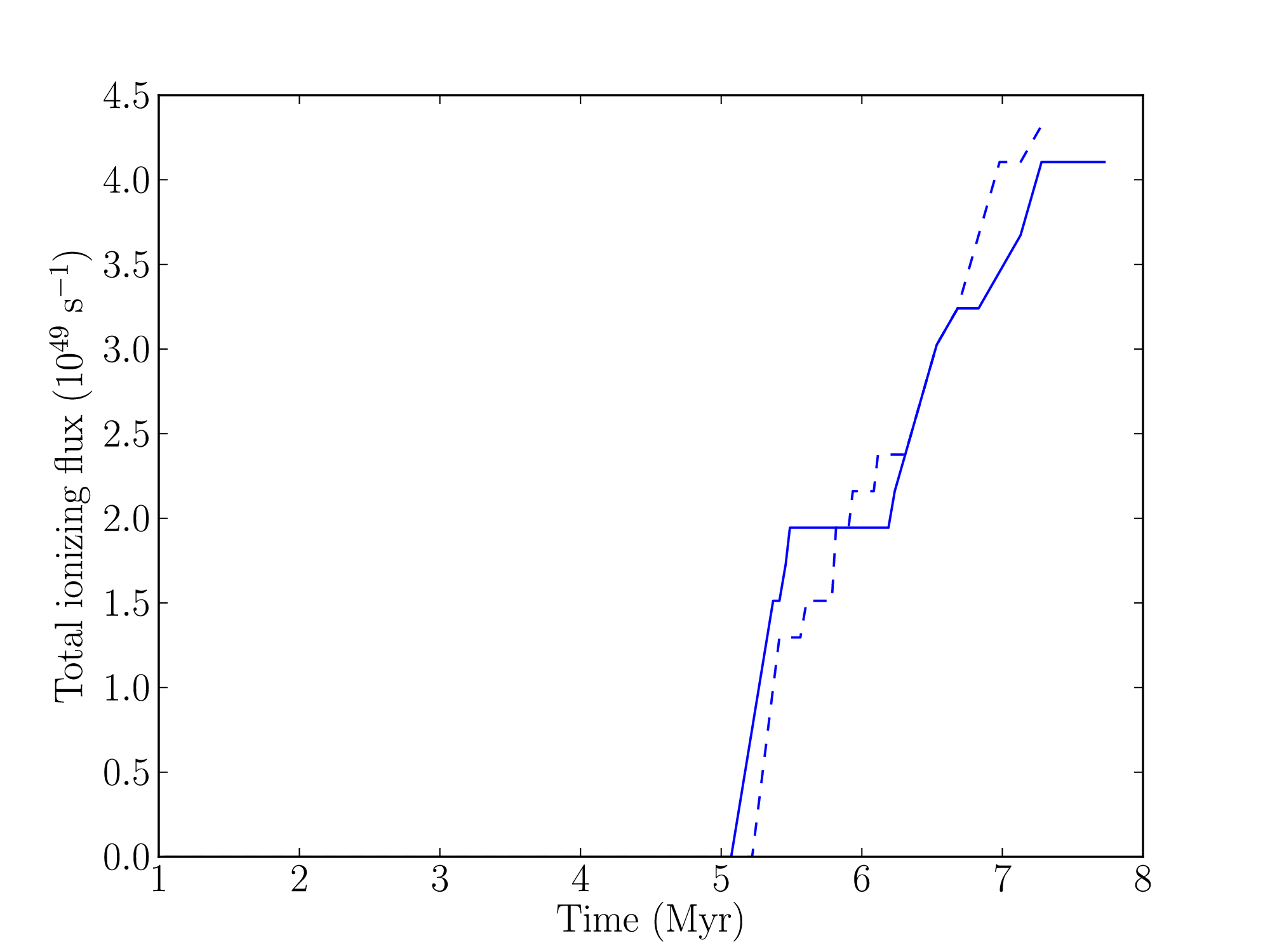}
\caption{Comparison of the total ionizing fluxes as functions of time in the standard (solid lines) and low--resolution (dashed lines) Run E.} 
\label{fig:rune_res_Q}
\end{figure}

To evaluate the influence of numerical resolution, we simply randomly dispensed with half the SPH particles in the initial conditions (reducing the particle number to 5$\times10^{5}$) for Run E and repeated the simulation. We show the results in Figure \ref{fig:rune_res} as dashed lines compared with the original $10^{6}$ particle calculation shown as solid lines. The lower resolution simulation has slightly lower fractions of ionized and unbound gas in the earlier stages of the simulation and slightly higher fractions in the later stages, although the stellar mass fractions are virtually identical. The reason for the differences is largely due to differences in the total ionizing luminosities in the two calculations, shown in Figure \ref{fig:rune_res_Q} -- the total luminosity in the low--resolution simulation is somewhat lower in the early phases of the calculation, but transitions to being the same as, or higher than, the luminosity in the standard run at later times. This is due simply to the slightly different history of cluster formation and mergers in the two simulations and illustrates how small--number statistics of the few ionizing sources can have some influence on the outcome of simulations, but the difference in the results is not large enough to be significant.

\label{lastpage}

\end{document}